\documentclass[10pt]{article}
\usepackage[margin=0.6in]{geometry}
\usepackage{hyperref}
\usepackage{multicol}
\usepackage{blindtext}
\usepackage{amsmath}
\usepackage{amssymb}
\usepackage{amsthm}
\usepackage{mathrsfs}
\usepackage{dsfont}
\usepackage{framed}
\usepackage{algorithm}
\usepackage{algpseudocode}
\usepackage{authblk}
\usepackage{graphicx}
\usepackage{wrapfig}
\usepackage[table]{xcolor}
\usepackage{tabularx}
\usepackage{array}
\newcommand{\whitecell}{\cellcolor{white!}} 
\usepackage{wrapfig}
\usepackage{float}
\usepackage{mdframed}
\usepackage{multirow}
\usepackage{booktabs}
\usepackage[table]{xcolor} 
\mdfdefinestyle{mddefault}{
rightline=true,innerleftmargin=4pt,innerrightmargin=4pt,innertopmargin=4pt,innerbottommargin=2pt
frametitlerule=false,backgroundcolor=black!5}

\usepackage{framed,color}
\definecolor{shadecolor}{rgb}{0.85,0.85,0.85}
\usepackage{ulem}
\usepackage{enumitem}
\usepackage{url}
\usepackage{cleveref}
\usepackage{hyperref,xcolor}
\usepackage{siunitx}
\hypersetup{
colorlinks,
linkcolor={red!50!black},
citecolor={blue!50!black},
urlcolor={blue!80!black}
}
\usepackage{xr}

\usepackage{tikz}
\makeatletter
\newcommand*{\addFileDependency}[1]{
\typeout{(#1)}
\@addtofilelist{#1}
\IfFileExists{#1}{}{\typeout{No file #1.}}
}\makeatother

\usepackage{lineno}
\usepackage{color}
\usepackage{ulem}
\usepackage{cancel}

\RequirePackage{tikz} 
\newcommand{\orcidicon}{\includegraphics[width=0.32cm]{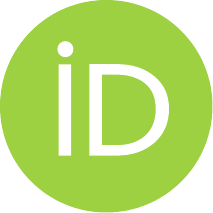}}

\foreach \x in {A, ..., Z}{%
\expandafter\xdef\csname orcid\x\endcsname{\noexpand\href{https://orcid.org/\csname orcidauthor\x\endcsname}{\noexpand\orcidicon}}
}

\title{Do Generalized-Gamma Scale Mixtures of Normals Fit Large Image Datasets?}
\author{Brandon Marks $^{1a}$, Yash Dave $^{1b}$, Zixun Wang $^2$, Hannah Chung $^2$, Riya Patwa $^2$, Simon Cha $^2$, Michael Murphy $^2$, Alexander Strang $^{2+}$ \orcidA{} }

\affil{\footnotesize $^{1a}$ Stanford University. Statistics \quad $^{1b}$ Stanford University. Institute for Computational and Mathematical Engineering \\
$^{2}$ University of California, Berkeley. Statistics \quad
$^{+}$ Correspondence: alexstrang@berkeley.edu}

\begin{document}

\maketitle

\tableofcontents

\pagebreak 


\section{Importance Statement} 

Ill-conditioned inverse problems occur frequently in imaging. In an ill-conditioned problem, estimation based on likelihood alone is unstable and imprecise. Bayesian approaches introduce prior information to bias inference towards solutions that are more likely under the prior. The realism of the posterior probabilities used to guide inference and to quantify uncertainty depends on the realism of the prior. In this paper, we test the realism of a family of parametric priors proposed for quasi-sparse Bayesian inference. While this family of priors has attracted interest in the computational imaging community, there is little work demonstrating their fit to real data. We offer the first empirical evidence for their realism across multiple large datasets drawn from different imaging fields. 

\section{Abstract} 

A scale mixture of normals is a distribution formed by mixing normal distributions with fixed mean but different variances. A generalized gamma scale mixture draws the variances from a generalized gamma distribution. Generalized gamma scale mixtures of normals are an attractive class of parametric priors for Bayesian inference in inverse imaging problems. Generalized gamma scale mixtures have two shape parameters, one that controls the behavior of the distribution about its mode, and the other that controls its tail decay. 
In this paper, we provide the first demonstration that the prior model is realistic for multiple large imaging datasets. We draw data from remote sensing, medical imaging, and image classification applications. We study the realism of the prior when applied to Fourier and wavelet (Haar and Gabor) transformations of the images, as well as to the coefficients produced by convolving the images against the filters used in the first layer of AlexNet, a popular convolutional neural network trained for image classification.  We discuss data augmentation procedures, procedures for identifying approximately exchangeable coefficients, and characterize the parameter regions that best describe the observed datasets. These regions are significantly broader than the region of primary focus in computational work. We show that this prior family provides a substantially better fit to each dataset than any of the standard priors it contains. These include Gaussian, Laplace, $\ell_p$, and Student's $t$ priors. Finally, we identify cases where the prior is unrealistic and highlight characteristic features of images that suggest the model will fit poorly.


\section{Introduction and Background}

\subsection{Background}

Inverse problems in imaging are often ill-posed or ill-conditioned. In either case, many solutions could correspond to the same or similar data, so the likelihood alone allows too many plausible solutions to direct useful recovery. To produce useful inferences, prior assumptions may be introduced to constrain conclusions. 

The Bayesian approach, which models the parameter and observation vectors as jointly random, encodes beliefs about the unknown signal through a prior. The prior biases inferences while reducing their variance across datasets. If properly calibrated, the error induced by bias is offset by the reduced variance per estimate, reducing the expected error across repeated applications of the inferential pipeline. A joint probability model for the signal and the data also promises self-consistent, conceptually straightforward uncertainty quantification (UQ) since, in the Bayesian setting, the full inferential solution is a posterior distribution that compromises between fidelity to the data and similarity to the prior. 

Extending the probability model to cover the unknown signal requires a prior. Constructing well-justified priors is often challenging, so Bayesian methods often suffer from misspecification. While the plausibility of a model may be validated using posterior predictive checking or cross-validation, its fit may be improved via empirical Bayes \cite{gelman1995bayesian}. While the sensitivity of associated inferences may be studied by varying prior beliefs \cite{si2024path}, the need for well-justified, if not well-specified, priors remains. 

Many practitioners adopt the assumption that real signals are sparse, compressive, or close to sparse in some shared representation \cite{candes2006stable,donoho2005stable,figueiredo2007gradient}. That is, most of the entries of a collection of signals are equal to or close to zero in some basis. Traditionally, sparsity is promoted by solving for point estimates that optimize a regularized loss function:
\begin{equation}
\label{eqn:penalty}
F_p(x; \lambda)= L(x,b)+\lambda\|x\|_p^p {,}
\end{equation}
where $b$ is the observed signal, $L$ is a log likelihood,  $\lambda$ is the regularization parameter that determines the magnitude of the penalty term, and $p$ is a shape parameter that determines solution properties. 

Recast in a Bayesian framework, minima of Equation \ref{eqn:penalty} are maximum a posteriori (MAP) estimates with a prior proportional to $\exp(-\nu \|x\|_p^p)$ for some $\nu$. Other sparsity-promoting priors abound, including Laplace \cite{babacan2009laplace, figueiredo2007laplace}, Cauchy \cite{markannen2019cauchy, suuronen2022cauchy}, and Horseshoe \cite{carvalho2009horseshoe, dong2023horseshoe, uribe2023horseshoe} priors. While the global shape of the prior may not matter for the MAP solution, it does influence the posterior probabilities used to quantify uncertainty. To pursue Bayesian UQ, priors should offer calibrated probabilities and should not be selected for their regularizing influence alone.





\subsection{The Prior Model}
\label{section:prior model}

\begin{figure}
    \begin{center}
  \begin{minipage}{1\textwidth}
     \centering
     \includegraphics[width=.7\linewidth]{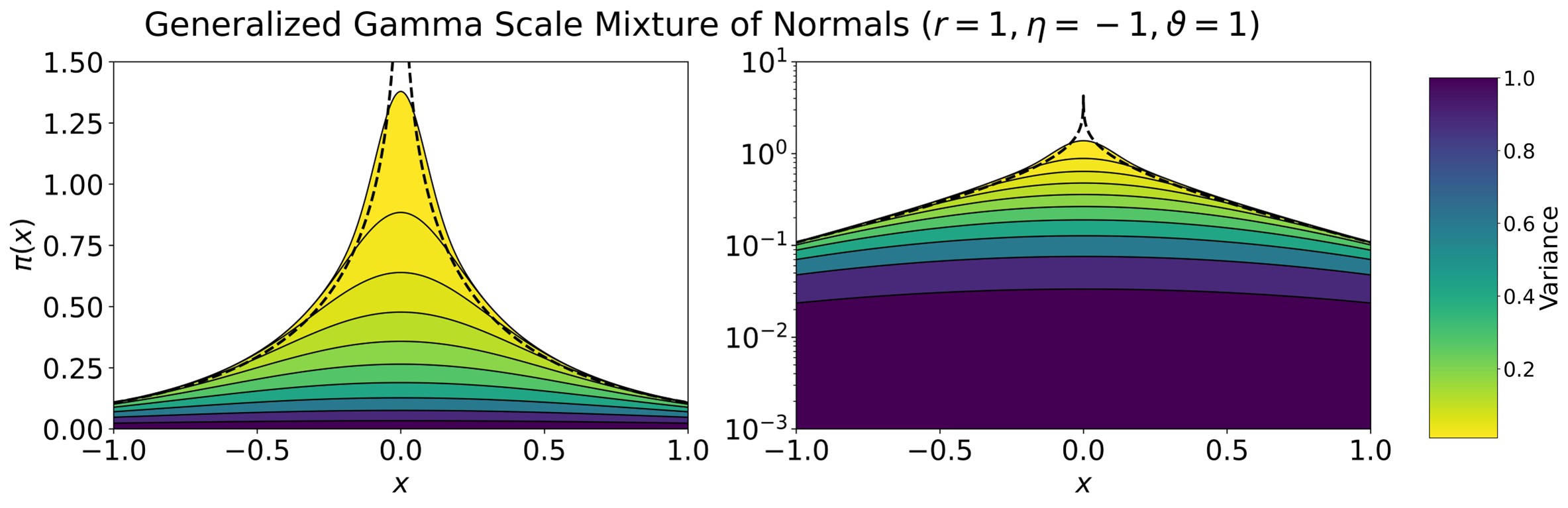}
  \end{minipage}\hfill
  \end{center}
  \caption{Each layer of the stacked plot is a normal distribution with mean 0 and variance drawn from a generalized gamma distribution with parameters ($r=1, \eta=-1, \vartheta=1$). The colors represent the values of the variance of that particular normal distribution. Stacking the 10 distributions together produces a discrete approximation to the integral in equation \eqref{eqn:prior}. The dotted lines indicate the true distribution. Here, the scale mixture of normals produces a distribution with a kink at 0 (the density is not continuously differentiable) and with exponential tail decay (slower than Gaussian).}
\label{fig:mixture_model}
\end{figure}

In this paper, we study generalized-gamma scale mixtures of normals. A scale mixture of normals is a distribution formed by mixing a collection of normal distributions with fixed mean but different variances. Scale mixtures of normals are a reasonably flexible family of even, unimodal distributions extensively used in hierarchical models for efficient robust inference \cite{abanto2010robust, calvetti2020sparse, calvetti2019hierachical,gneiting1997normal}. 
Any univariate random variable $X$, with density $f_X$ that is symmetric about zero and is completely monotone, i.~e.~$(-\frac{d}{dy})^k f_X(\sqrt{y}) \geq 0$ for all $k \in \mathbb{Z}$ and all $y \geq 0$, may be expressed as a scale mixture of normal random variables \cite{andrews1974}.
We fix the mean to zero, and draw the variances from a generalized gamma distribution (Fig. \ref{fig:mixture_model}). 

All mixture models can be expanded hierarchically. Hierarchical models are attractive in large-scale Bayesian inference since their architecture allows efficient coordinate-based computation and organized information synthesis. Both features are attractive in imaging. Efficient methods are essential since images are high-dimensional. Hierarchical models can also synthesize global and local information by resolving separate regularizers at different locations in an image. Examples include the Horseshoe prior. Our model follows this conceptual tradition. See \cite{jassal2025local} for similar approaches.

Recast hierarchically, we study a prior where:
\begin{equation}
x_j|\theta_j \sim \mathcal{N}(0, \theta_j), \text{ for } \theta_j >0, 1\leq j \leq n
\end{equation}
and where $\theta_j$ denotes an unknown prior variance. The variances $\theta$ are drawn independently from a generalized gamma distribution with parameters $r_j$, $\eta_j$, and $\vartheta_j$:
\begin{equation}
    \pi_{\text{hyper}}(\theta_j|r_j,\eta_j,\vartheta_j) = \prod_{j=1}^n \frac{|r_j|}{\Gamma(\beta_j)}  \frac{1}{\vartheta_j} \left( \frac{\theta_j}{\vartheta_j} \right)^{\eta_j + \frac{1}{2}} \exp \left(-\left( \frac{\theta_j}{\vartheta_j} \right)^{r_j} \right),
\end{equation}
where $\beta_j = (\eta_j + 3/2)/r_j$ for $\eta_j$ such that $\beta_j > 0$ \cite{calvetti2009conditionally,calvetti2020sparse}. 

The hyperparameters $r_j \in \mathbb{R} \setminus \{0\}$ and $\eta_j$ are shape parameters, while $\vartheta_j > 0$ are scale parameters. Marginalizing over $\theta_j$ produces the mixture form for the prior model:

\begin{equation}
\label{eqn:prior}
    \pi(x_j|r_j,\eta_j,\vartheta_j) = \int_0^\infty \pi(x_j|\theta_j) \pi_{\text{hyper}}(\theta_j|r_j,\eta_j,\vartheta_j) d\theta_j
\end{equation}

Studying the univariate case ($n=1$) is sufficient since the variances are chosen independently. In principle, we could use separate parameters for each component, but to control model complexity, we group the components into blocks that share the same prior parameters. Subscripts denoting dimension are subsequently dropped. 
In the rest of this paper, the \textit{prior model} refers to this mixture parameterized by $r, \eta, \vartheta$, with the scale parameter set to $1$ unless otherwise specified. 

\begin{figure}[t!]
    \begin{center}
     \centering
     \includegraphics[width=0.95\linewidth]{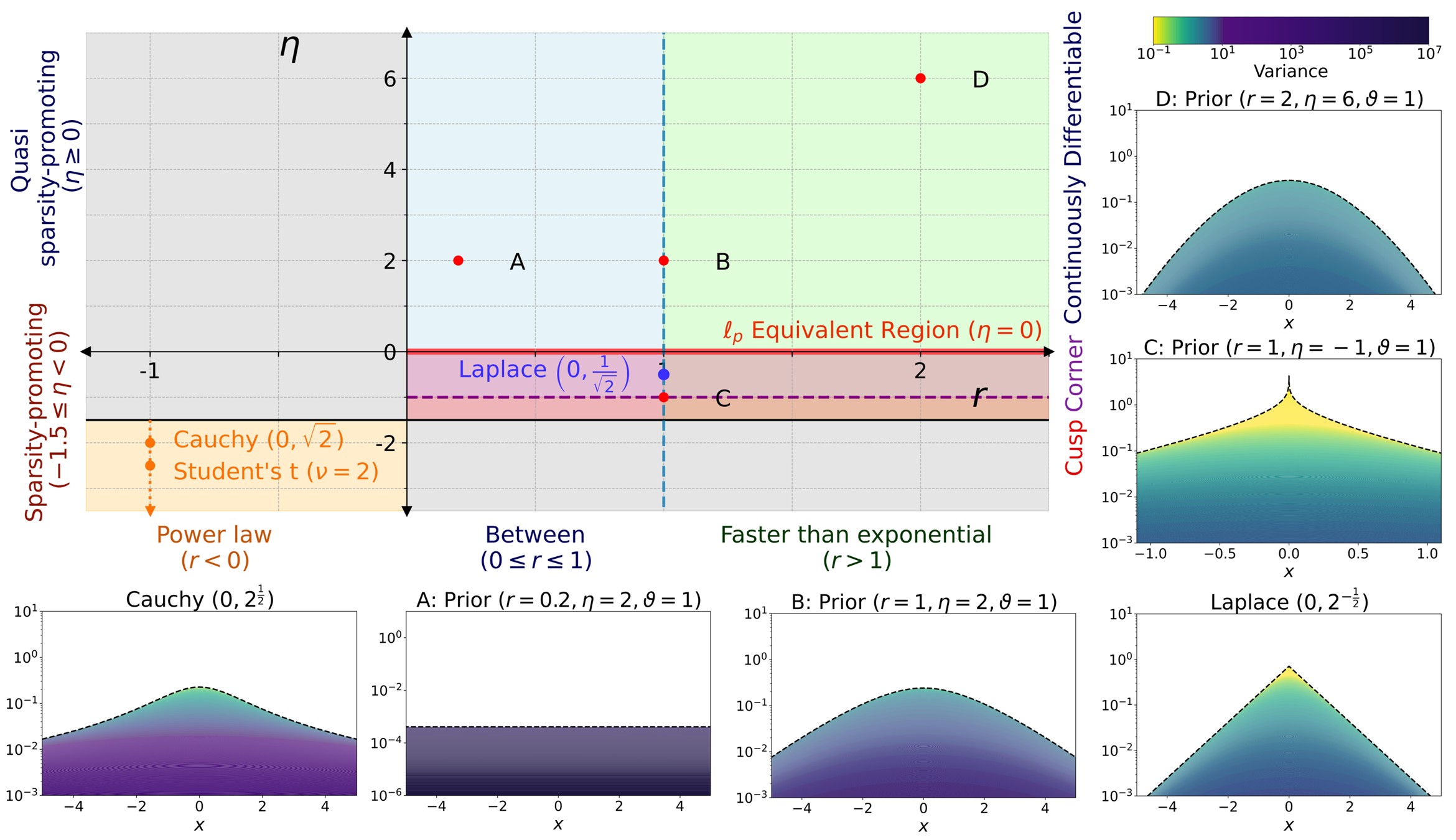}
  \end{center}
  \caption{Parameter space map with commonly used priors (Cauchy, Student's $t$, Laplace) labeled. The scale parameter $\vartheta$ is fixed to be $1$.}
\label{fig:parameter space map}
\end{figure}



This family of priors is moderately expressive. Unlike traditional priors, it offers separate shape control over its tail and peak. Fig. \ref{fig:parameter space map} provides examples for different values of $r$ and $\eta$. 

The shape parameter $\eta$ controls the peak behavior for positive values of $r$. The $\eta=0$ line separates the space into priors that strongly promote sparsity and those that promote quasi-sparsity in MAP estimators. Quasi-sparsity means that most of the entries of the signal are close to, but not exactly, zero. The peak of the distribution is smooth and continuously differentiable in these cases (see panel \textit{B}). For $\eta<0$, the distribution is non-differentiable at zero. The peak develops a corner as $\eta$ crosses $0$ from above (progress from panel \textit{B} to panel \textit{C}). As $\eta$ crosses $-1$ from above, the distribution develops a singularity at 0. It exhibits a cusp at 0 for all $\eta \leq -1$. 

The $\ell_p$-Equivalent Region ($\ell_pER$), marked in red, refers to the $r > 0, \eta=0$ half line. Setting $r > 0$ and $\eta = 0$ reproduces the commonly studied $\ell^p$-penalty prior \cite{calvetti2020sparse}, with $p = 2r/(1 + r)$.
%
%
Therefore, as $\eta$ tends to 0 from above, MAP estimates converge to the $\ell_p$ penalized estimates for $0<p<2$ \cite{calvetti2019brain, calvetti2019hierachical}.

The remaining shape parameter $r$ controls the tail decay rate of the prior. For $r<0$, the tail decay follows a power law.  In the log-density plot for the Cauchy, for example, note how the slope of the tails flattens as $|r|$ tends to zero.
Comparing the subplots of Figure \ref{fig:parameter space map} corresponding to points \textit{A} and \textit{B}, shows that increasing $r$ from $0.2$ to $1$ with fixed $\eta=2$ shifts the prior from a nearly uniform prior to one with exponential tail decay. The distributions labeled \textit{B} and \textit{C} both exhibit exponential tail decay as illustrated in the log-density plot. Finally, for $r>1$, the priors' tails decay faster than exponentials. As $r$ approaches infinity, the prior acquires Gaussian tails (c.f.~the log-density plot of panel \textit{D}).

Classical priors, including the Laplace, Cauchy, Student's $t$, and Gaussian, are special cases of the mixture family. These are also included in Figure \ref{fig:parameter space map}. For example, exponential mixing ($r=1$) produces a Laplace prior \cite{andrews1974}, with scale parameter a function of $\eta$ and $\vartheta$. The Cauchy and Student's $t$ are mixtures of normals with inverse gamma variances ($r=-1$). 

To obtain a Gaussian, observe that, as $r, \eta \to \infty$, the generalized gamma hyperprior concentrates. In turn, the hierarchical model collapses to a Gaussian prior. Limits following different level sets of the variance point to Gaussian priors with different standard deviations. Not all sparsity-promoting priors are included in this family of priors. Important examples include spike-and-slab priors, \cite{mitchell1988spike} Horseshoe priors \cite{carvalho2009horseshoe, dong2023horseshoe, uribe2023horseshoe}, and elastic-net priors \cite{zou2005elastic}.

\begin{figure}[h]
    \begin{center}
     \centering
     \includegraphics[width=.9\linewidth]{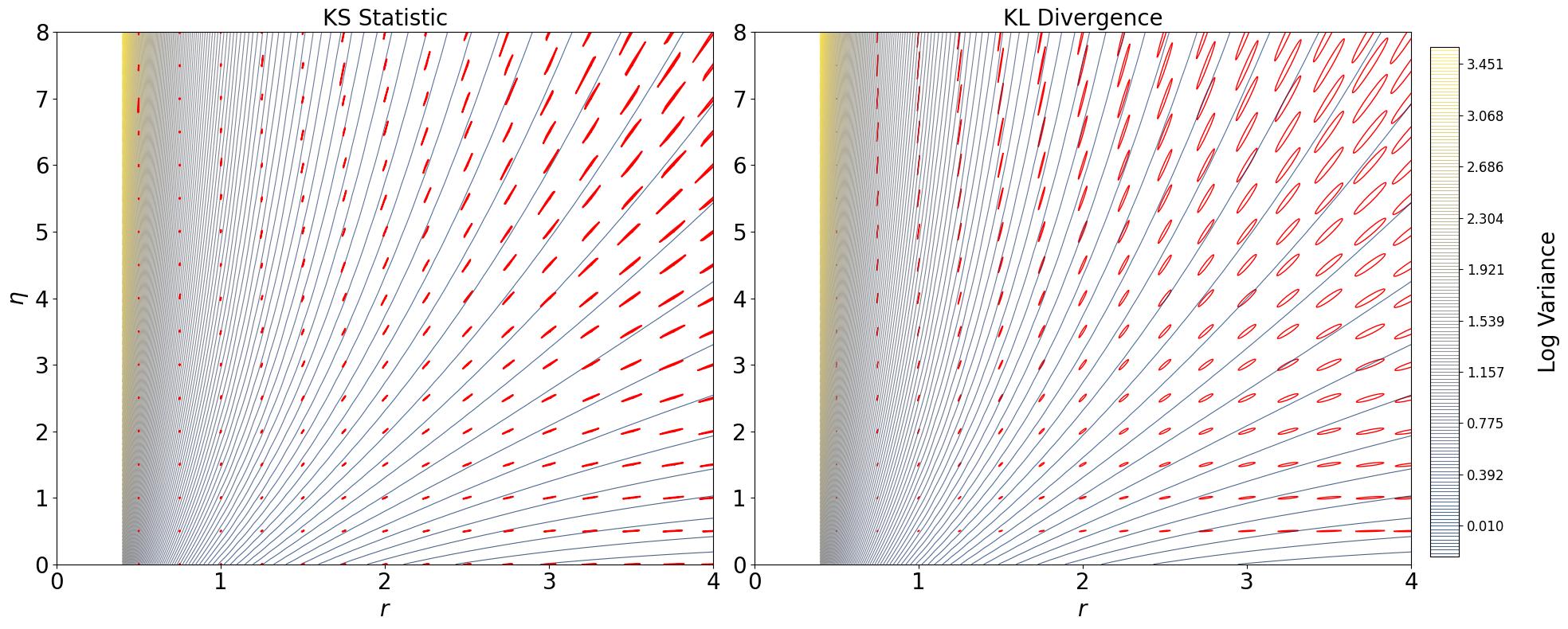}
  \end{center}
  \caption{Sensitivity of the prior to perturbations in the shape parameters $r, \eta$ as measured by the KS statistic and the KL divergence for fixed scale $\vartheta = 1$. The red regions in the left plot denote $\epsilon$-level sets of the KS statistic between a reference distribution, defined by the parameter pair at the center of each region, and the distribution produced by perturbing $r$ and $\eta$. The red ellipses on the right plot indicate $\epsilon$-level sets of the KL divergence. The background curves indicate the level sets of the prior variance. The prior is most sensitive to parameter changes that change its variance, as shown by the fact that the red level sets are oriented to allow larger variations parallel to level sets of the prior variance.}
\label{fig:ks kl plot}

\end{figure}

Figure \ref{fig:ks kl plot} illustrates the sensitivity of the prior distributions to perturbations in the shape parameters for a fixed scale. When divergences between distributions are measured using either a Kolmogorov-Smirnov (KS) distance, or a Kullback-Leibler (KL) divergence, the prior is most sensitive to changes in the shape parameters that change its variance. For a fixed scale, the shape parameters are reasonably identifiable.

\begin{figure}
    \begin{center}
     \centering
     \includegraphics[trim = 0 0 0 0, clip, width=.75\linewidth]{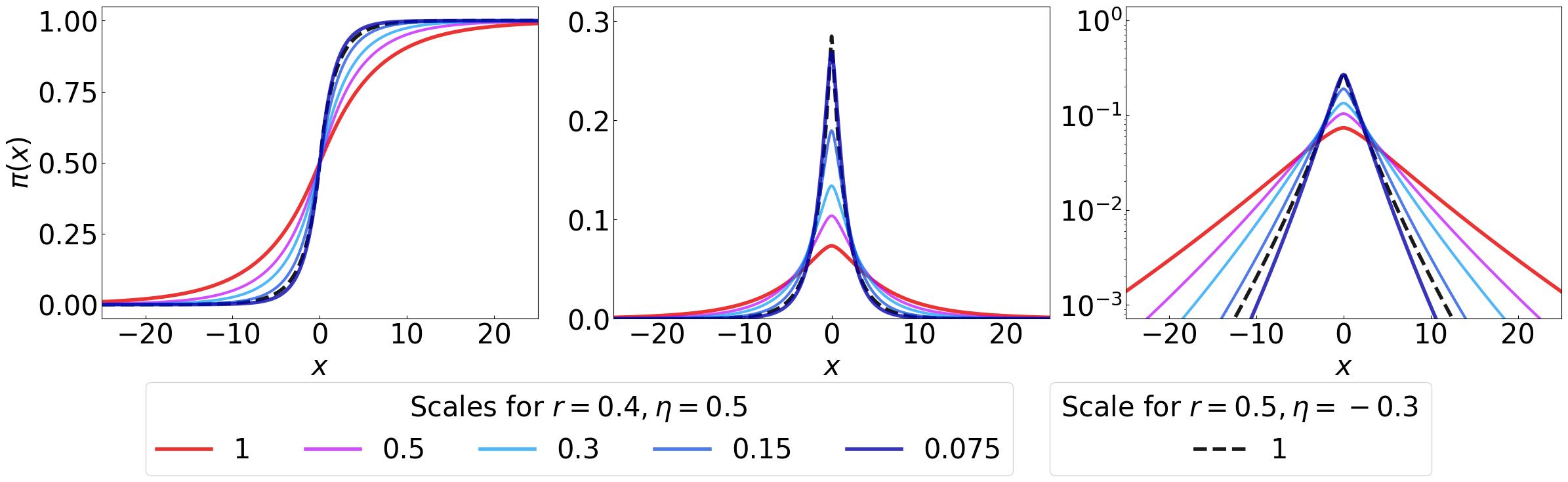}
  \end{center}
  \caption{Changes in the scale parameter $\vartheta$ can be used to compensate for changes in $r$ and $\eta$. The red and black curves differ in their tail decay rates and smoothness of their peaks. Decreasing $\vartheta$ bridges this gap, with $\vartheta = 0.075$ producing approximately the same distribution. The log-density plot shows that the tails are slightly smaller for the blue curve, but the peak of the distribution is a near-perfect match.}
  \label{fig:scale similarity}
\end{figure}

Adjusting the scale parameter, $\vartheta$, can stretch or compress the distribution. By choosing suitable values, distinct ($r$, $\eta$) pairs can be made to produce \textit{similar} distributions.  Figure \ref{fig:scale similarity} shows examples for $\eta=0.5$ and $\eta=-0.3$. 
As a result, we will struggle to identify parameter triples along one-dimensional manifolds where variations in the shape parameters can be approximately accounted for by a matching variation in scale. 

 
There is an extensive literature on numerical methods with this hierarchical prior.
Initially, Chantas et.~al.~introduced hierarchical spatially-adaptive Gaussian priors with gamma hyperpriors for image restoration via an iterative modified Newton algorithm \cite{chantas2006bayesian}.
Calvetti and Somersalo (2008) adopted a similar cyclic iteration method, and recast Total Variation and Perona–Malik regularization schemes within the Bayesian framework \cite{calvetti2008hypermodels}. Babacan et.~al.~(2010) proposed a similar scheme and demonstrated its correspondence to iteratively reweighted least squares. 
Expanding on the convexity analysis provided in \cite{calvetti2020sparse}, hybrid solvers that switch between locally and globally convex models were proposed in \cite{calvetti2020sparsity, si2024path}. Subsequent developments sought generalization to nonlinear inverse problems. Kim et.~al.~proposed a Kalman filtering variant \cite{kim2022hierarchical} and Manninen et.~al.~adapted the algorithm to a nonlinear problem from diffuse optical tomography \cite{manninen2024dot}.
In parallel, Glaubitz et.~al.~(2023) generalized existing methods to arbitrary linear transforms and forward models, providing algorithms to compute posterior approximations for generalized sparse priors beyond canonical bases \cite{glaubitz2023algorithm}. Most recently, attention has shifted from producing point estimates to full-scale posterior inference and uncertainty quantification (UQ).
Lindbloom et.~al.~generalize the algorithm to allow noninvertible sparsifying transforms and unknown noise variance \cite{lindbloom2025efficient}. 
Xiao and Glaubitz (2023) extend hierarchical learning to sequences of images, coupling hyperpriors temporally to improve sequential recovery \cite{xiao2023algorithm}.
Recent work by Calvetti et.~al.~(2024) and Glaubitz (2025) also studies sampling strategies from these sparsity-promoting hierarchical models \cite{calvetti2024sampling, glaubitz2025efficient}.

All of this work refines and generalizes inference methods conditional on the prior model. There is little work investigating whether the priors are plausible. 

\subsection{Motivation}
\label{section:motivation}

The priors used in Bayesian imaging often represent subjective beliefs about expected regularities of the unknown solution. Common assumptions include sparsity, smoothness, piecewise regularity, and self-similarity \cite{buades2005assumption, chatterjee2012assumption, orieux2010assumption, pan2013assumption,  roininen2019assumption}. The prior model studied supports these assumptions. These assumption-driven priors contrast modern image priors, which are increasingly data-driven and use machine learning techniques to extract and encode prior knowledge \cite{ gonzalez2022learned, holden2022learned, mukherjee2022learned, thong2024empirical}.  

To our knowledge, our paper describes the first empirical attempt to evaluate the realism of generalized gamma scale mixtures of normals specified in Equation \eqref{eqn:prior}. We surveyed 18 papers that use the prior model spanning 2008-2025 to understand previous validation efforts. For details, see Appendix \ref{app:survey}.
We find that 16 (78\%) papers use synthetic or computed examples to demonstrate the efficacy of their methods. Of the remaining four papers, two use select standard images (e.g.~the Shepp-Logan Phantom) and rely on qualitative checks for validation. 
The other two papers use a combination of synthetic, standard, and real data \cite{calvetti2019brain, xiao2023algorithm}. In \cite{calvetti2019brain}, the sensitivity and specificity of the inverse solver is systematically tested using synthetic data. The real data, augmented with realistic activity in the brain stem, is used to measure how much, on average, the reconstructed activity is concentrated in the brain region of the simulated active patch. 
Conversely, in \cite{xiao2023algorithm}, in addition to MRI phantom images, three images from a real road-traffic monitoring dataset are used to demonstrate the qualitative performance of the proposed joint recovery method compared to separate recovery methods. Importantly, none of the papers validate the prior model in isolation. The scarcity of studies validating the prior family motivates the following question: 

\vspace{0.05 in}
\textbf{Does the prior model fit empirical data?}
\begin{enumerate}[label=(\alph*), leftmargin=1.2cm]
    \item If so, for what datasets?
    \item Under which representations?
    \item And for which parameters?
\end{enumerate}

We evaluate the fit of the prior model by examining two main assumptions. First, independence of entries after a change of basis. Second, the parametric form of the marginals. We test the latter in detail, and perform exploratory tests to study the independence assumption.

\subsection{Outline}

 The datasets used are described in Section \ref{section:data}. In Section \ref{section:methods}, we detail the entire testing pipeline used. In particular, Section \ref{section:normalization} describes data augmentation procedures and their relation to downstream inference tasks. The sparsifying transforms applied to the images are discussed in Section \ref{section:representation}, namely, the Fourier transform, Gabor and Haar wavelet transforms, and learned filters from the first layer of AlexNet. The remainder of Section \ref{section:methods} discusses the metric by which we define a good fit, the procedure by which we find best-fit parameters, and additional hypothesis tests we use. For each dataset, we report our findings in Section \ref{section:results} in response to the motivating questions. In Section \ref{section:discussion}, we discuss our findings broadly and outline the limitations of our study.



\section{Data}
\label{section:data}

We study three categories of image data: remote sensing, natural, and medical. 

\begin{table}[h]
\caption{Summary of remote sensing (agriVision, pastis, spaceNet), natural (coco, segmentAnything), and medical (syntheticMRI2D, syntheticMRI3D) image datasets used.}
\centering
\renewcommand{\arraystretch}{1.5}
\begin{tabular}{@{}ccp{9.7cm}}
\toprule
\textbf{Name} & \textbf{\# Images} & \textbf{Description} \\ \midrule

agriVision & 4500 & Agricultural Vision, central US farmlands ($256 \times 256$) \cite{agriVision} \\
pastis & 1590 & Panoptic Agricultural Satellite Time Series, farm fields in France ($128 \times 128$) \cite{pastis} \\
spaceNet & 3401 & Multi-Sensor All Weather Mapping Dataset, centered in Rotterdam ($400 \times 400$) \cite{spaceNet} \\ \midrule
coco & 4050 & Common Objects in Context, split into indoor/outdoor subsets ($256 \times 256$) \cite{coco} \\
segmentAnything & 7072 & Natural image dataset, compiled for training segmentation models ($512 \times 512$) \cite{segmentAnything} \\ \midrule
syntheticMRI2D & 15000 & 2D synthetic MRI brain images, split into axial ($333 \times 234$), coronal ($244 \times 234$), and  sagittal slices ($174 \times 234$) \cite{syntheticMRI} \\
syntheticMRI3D & 300 & 3D synthetic MRI brain images ($160 \times 224 \times 160$) \cite{syntheticMRI} \\ \bottomrule
\end{tabular}
\label{tab: datasets}
\end{table}

\begin{figure}
   \centering
    \includegraphics[width=0.8\linewidth]{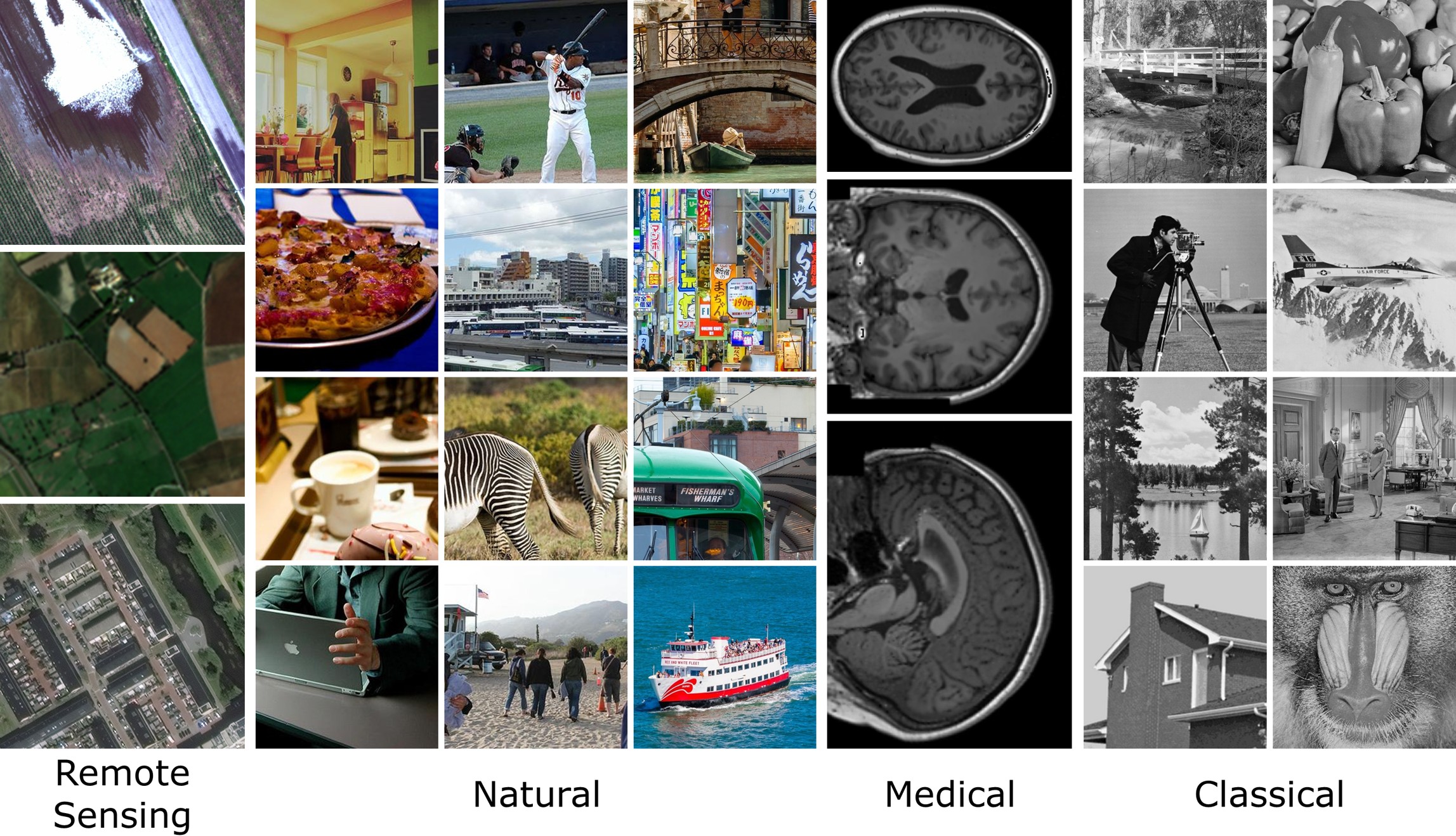}
  \caption{Representative sample(s) from the remote sensing, natural, medical, and classical image datasets.}
\label{fig:data panel}
\end{figure}

\subsection{Remote Sensing}

The images comprise aerial shots of farm fields with various crops, lakes, trees, roads, buildings, and cars. The limited color palette and the strongly defined edges formed by tract bounds and roads suggest sparsity in Fourier and Haar wavelet representations. 

\subsection{Natural Image Sets}

Natural images include a greater diversity of colors, contrasts, and objects than the remote sensing datasets. These images have been used to train contemporary machine learning models on various tasks, including segmentation, object detection, and image captioning. The images are clustered by subject. Clusters pose a challenge for the prior model, which assumes unimodality in each component. We use natural image datasets to test whether images clustered by subject can be modeled using a unimodal family if represented using a basis that focuses on textural or global features.

For comparison, we collected a set of nine classical images commonly used in the image processing literature (classical panel of Figure \ref{fig:data panel}). We use this set to test whether this small set of frequently referenced images are, or are not, representative of large image datasets.

\subsection{Medical Imaging}

These high-resolution brain images are synthetically generated using a Latent Diffusion Model trained on UK Biobank data. See \cite{syntheticMRI} for details about the methodology and veracity of the generated images. Previous work with generalized gamma scale mixtures of normals has included applications to medical domains like electroencephalography and diffuse optical tomography \cite{calvetti2009conditionally, calvetti2019brain, manninen2024dot}. These examples motivate an empirical validation study for brain images. We use the 3D brain images to explore how sparsity assumptions extend to higher dimensions.



\section{Methods}
\label{section:methods}

All code and data used in this study are publicly available in \href{https://github.com/yashdave003/gengamma-scale-mixture-validation-images}{this repository}.


\subsection{Data Treatment}
\label{section:data treatment}

Before testing, we convert our data into collections of coefficients by changing the image representation or by passing through a filter bank. Sections \ref{section:normalization}, \ref{section:representation}, \ref{section:grouping via approximate exchangeability}, and \ref{section:grouping_via_induced_exchangeability} outline the procedures used to standardize, transform, and then group the coefficients. Each step imitates a preprocessing or modeling procedure plausibly used in applications. Here, we provide a conceptual framework that organizes the different arguments used to justify our data treatment.

Since we apply our hypothesis tests to the treated data, each data treatment step encodes a modeling assumption. Accordingly, a data treatment step should reasonably imitate actual data treatment in inverse imaging while corresponding to a justifiable modeling assumption. Since modeling assumptions constrain the space of available models, each assumption introduces different misspecification errors. So, to justify a data treatment step, we need to understand the errors it will produce. We need a contextual categorization of errors that distinguishes acceptable and unacceptable errors. 

The priors tested in this paper were proposed for inverse imaging problems, primarily in medicine (c.f.~\cite{calvetti2005localreg,calvetti2009conditionally,calvetti2019brain}). 
In a Bayesian inverse imaging context, a practitioner uses a prior, a likelihood model, and the observed data to form a posterior, draws inferences from summaries of the posterior, then makes decisions based on those inferences. 
Errors in the prior are only problematic if they induce errors in inference that confound decision-making. 
This perspective suggests the following categorization of acceptable misspecification errors. See Appendix \ref{app:misspec_error_types} for further justification. 
A small error in the prior will induce a small error in the reconstruction, leading to a \textbf{Negligible} error in inference. 
Some errors in the prior may be larger but may not significantly affect the overall image reconstruction, provided that the information lost can be recovered through the likelihood. We categorize such errors as \textbf{Likelihood-Constrained}. 
Other model errors may result in substantial reconstruction errors but may not impact the inference itself; we label these as \textbf{Irrelevant Artifacts}. 
Finally, we classify errors as \textbf{Resolvable Artifacts} when they cause large reconstruction errors that could affect inference, but those artifacts can be easily corrected.

\subsubsection{Standardization}
\label{section:normalization}

To match generic image processing pipelines, we standardize each image by centering and scaling. Because the remote sensing images were obtained from larger spatial patches, standardization occurred on the patch level. 

Standardization produces better fits to the prior model since the prior assumes that all coefficient distributions are centered about zero. Scaling reduces the variation in exposure across images. We allow these transformations because any centering and scaling operation can be inverted by a linear transformation of the recovered image. Thus, standardization is justified by argument (4) in the error classification of Section \ref{section:data treatment}.
\subsubsection{Representation}
\label{section:representation}

The choice of transform is a modeling assumption since the prior model assumes both (1) independence between coefficients after the transformation, and (2) a parametric form for the coefficients' marginals. Because the prior family was introduced to model quasi-sparsity, we implicitly assume that (3) the coefficients produced by the transform have quasi-sparse marginals. 

A priori, we do not know which transform is likely to satisfy these assumptions for any ensemble of images. The transforms we considered were the Fourier transform, Haar and Gabor wavelets, and learned filters from AlexNet. These transforms represent different types of filter banks commonly applied to images, and often assumed to promote sparsity. 

\begin{figure}
    \begin{center}
     \centering
     \includegraphics[width=.95\linewidth]{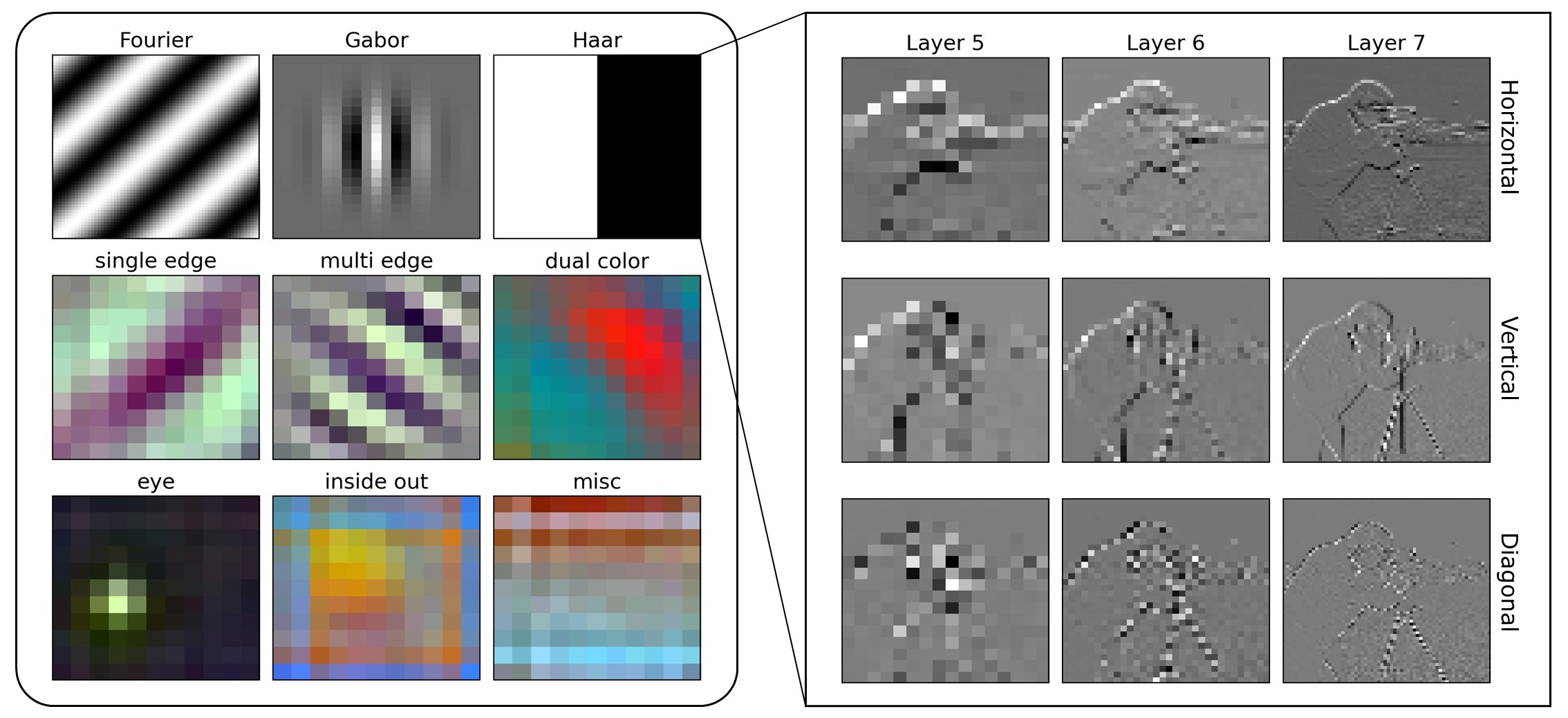}
  \end{center}
  \caption{Left panel: Visualization of representative filters for each transform (Fourier, Gabor, Haar, Learned) and filter group, where applicable. Learned filters are normalized 0-1 and extracted from the first layer of AlexNet. See Appendix \ref{app:learned_gabor_filters} for all 64 learned filters and 42 Gabor filters used. Right Panel: Example of Haar wavelet transform applied to the MIT cameraman image, ordered from low (Layer 5) to high (Layer 7) frequency components, for each orientation.}
\label{fig:filters_and_wavelet_layers}
\end{figure}

The \textbf{Fourier} transform is a common choice since it efficiently converts functions on the real spatial/pixel domain to functions on the complex frequency domain. We also test two classes of wavelet transforms that retain both spatial and frequency information. \textbf{Gabor} filters are defined by a sinusoidal carrier modulated by a Gaussian envelope. They comprise a smooth, infinitely differentiable wavelet basis and are well-suited to capturing oriented textures and mid-level features. In contrast, the \textbf{Haar} wavelet basis is piecewise constant and more effective at representing local patterns, particularly edges.
Figure \ref{fig:filters_and_wavelet_layers} shows a visualization of the different transforms. 
Finally, the 64 \textbf{learned} filters in the first layer of AlexNet were chosen for their utility in image processing tasks. For all of these transforms, satisfying the independence assumption means the features represented by the basis functions are independent of one another. 
For example, AlexNet coefficients corresponding to distinct parts of the images, or to filters that detect very different features, are plausibly independent, while filters corresponding to nearby parts may not be. 
Assuming the quasi-sparsity assumption means a majority of the image can be reconstructed from a few key basis vectors. 
For example, satellite images of farmed fields may be quasi-sparse under the Fourier transform because the images exhibit large patches of periodic behavior due to the field's structure.

\subsubsection{Grouping Approximately Exchangeable Coefficients}

\label{section:grouping via approximate exchangeability}


Fitting a distinct prior for each coefficient requires as many prior fits as there are pixels in the image. This would typically require three parameters for over $2^{16}$ distributions (see Table \ref{tab: datasets}), which is unrealistic. Therefore, it is necessary to group coefficients into blocks. Grouping introduces additional model assumptions. To justify grouping, we classify the ensuing misspecification errors via the categories provided in \ref{section:data treatment}. A summary of all grouping actions and associated error types can be found in Table \ref{tab: image_data_augment_mapping}, see Appendix \ref{app:grouping_args} for more detailed justifications.

Because we treat the coefficients as identical within each block, and assume that all coefficients are independent, grouping the coefficients is equivalent to assuming that all of the coefficients within a block are independent and identically distributed (i.i.d.). To isolate grouping arguments from the independence assumption, we assume first that the coefficients within a block are exchangeable, that is, statistically indistinguishable from one another, then, in addition, that they are independent. We test exchangeability and independence separately.

A sequence of random variables $\{X_j\}_{j=1}^n$ is exchangeable if its joint distribution is invariant to any permutation of the indices.
Coefficients that are approximately exchangeable form easily justified groups. A block of coefficients is approximately exchangeable if all of the images generated by swapping the blocked coefficients are close to equally likely under the data-generating process.  For example, Rotating an image 90$^{\circ}$ corresponds to swaps in the horizontal and vertical details after applying the Haar wavelet transform. If the probability of sampling an image and its 90$^{\circ}$ rotation are approximately the same, then the coefficients that swap roles under the rotation are approximately exchangeable. Grouping approximately exchangeable coefficients produces \textbf{negligible} errors  (see Section \ref{section:data treatment}).

We test the approximate exchangeability of two groups of random variables by examining how similar their distributions are. We adopt the two-sample Kolmogorov-Smirnov statistic to quantify the difference between empirical distributions. 

\subsubsection{Grouping Coefficients to Induce Exchangeability}

\label{section:grouping_via_induced_exchangeability}


Coefficients may be grouped to induce exchangeability. Any procedure that intentionally induces exchangeability discards information by treating distinguishable units as if they were indistinguishable. If the blocks are not approximately exchangeable, then ensuing errors may not be \textbf{negligible}, but may still fall into one of the other error types outlined in \ref{section:data treatment}. To justify the associated misspecification errors, we develop an equivalence between coefficient blocks and augmentation procedures.


Consider an image and an augmentation procedure that randomly permutes coefficients of the image within the same group. 
The coefficients of the augmented image are exchangeable within their groups. 
A prior model that uses one marginal for all the coefficients in a particular group, set to the average marginal over the group, assigns the correct marginal to the coefficients of the augmented image. 
Thus, we can map a grouping procedure to an augmentation procedure, then treat the grouped model as model for the output of the augmentation.

The augmentation procedure described above acts on the coefficients of the image after a change of representation. Each coefficient augmentation is equivalent to an augmentation applied to the original image. 
It is often easier to interpret the image augmentation. Both coefficient and image augmentation are impossible in an inverse problem, since the user does not have access to the true image. 
Instead, the user is given a noisy transformation of the image, pushed through a forward map. 
Pushing the action of an image augmentation through the forward map returns an equivalent data augmentation. The specific data augmentation depends on the forward map, and should change both the data and the noise model used to form a likelihood. 
Many forward models in imaging admit a symmetry of action, so commute with particular transforms. 
For instance, the forward models used in tomography commute with translations, rotations, and spatial scaling of the ground truth.

\begin{figure}[t!]
    \begin{center}
     \centering
     \includegraphics[width=.85\linewidth]{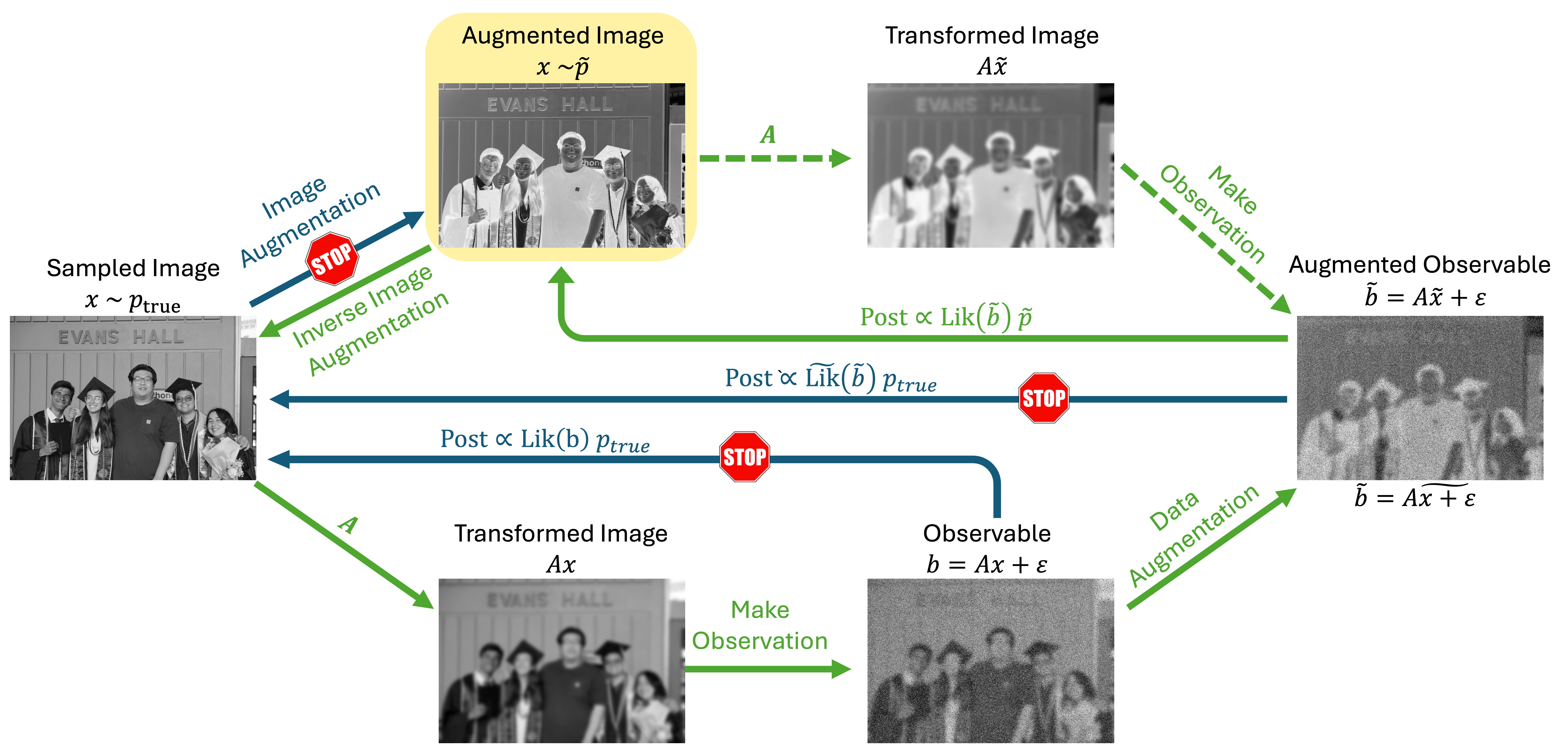}
  \end{center}
  \caption{Data augmentation model used to justify exchangeability. The solid green arrows represent the workflow in Section \ref{section:grouping_via_induced_exchangeability}. This pipeline aims to recover images specified under the augmented prior (highlighted in yellow). To recover the true images and perform inference, we propose the methods described in Section \ref{section:data treatment}, \textbf{resolvable artifacts}. The blue arrows with stop signs represent steps that are theoretically possible, but practically impossible as they involve (1) drawing an image from the augmented prior or (2) using the true prior $p_\text{true}$ to compute a posterior distribution. We do not attempt to model $p_{\text{true}}$. The green dashed arrows are steps that are both theoretically and practically possible, but that follow an impossible step.}
\label{fig:augmentation_graphic}
\end{figure}

Figure \ref{fig:augmentation_graphic} shows how discarding information in the model can be related to an augmented pipeline. Ideally, images are drawn from a true prior, $p_{\text{true}}$, passed through a forward map that models the imaging process, and then observed with measurement noise. The forward process is reversed using the true likelihood and prior. Since grouping is required, the true prior cannot be modeled perfectly. Instead, the simplified prior $\tilde{p}$ models the coefficient distribution after augmenting the images. This procedure is impossible in practice, so we consider the equivalent pipeline where the data is augmented, then inverted with the true likelihood.

These observations suggest the following program: (1) propose a group of coefficients, (2) express the grouping assumption as a prior model for the coefficients after an augmentation, (3) convert the coefficient augmentation to an image augmentation, (4) when the image augmentation commutes with the forward model, convert from an image augmentation to a data augmentation. Accept the group assumption if a full Bayesian pipeline, which absorbs the data augmentation into its likelihood, could reliably recover images drawn from the true prior if given unaugmented data. This is possible if (a) applying the true likelihood to unaugmented data would rule out augmented images that are possible under the image augmented prior, (b) the image augmentation is easily corrected given augmented images, or (c) recovery of an augmented image does not hinder downstream interpretation or decision-making. These cases correspond to the last three error categories of Section \ref{section:data treatment}. 

\begin{table}
\caption{Image augmentation methods,  coefficient groups, and misspecification error types as described in Table \ref{tab: fit categorization}.}
\centering
\renewcommand{\arraystretch}{1.5}
\begin{tabular}{
  >{\centering\arraybackslash}m{2.6cm} 
  >{\centering\arraybackslash}m{1.9cm} 
  >{\centering\arraybackslash}m{4cm} 
  >{\centering\arraybackslash}m{5cm}}
\toprule
\textbf{Image Augmentation} & \textbf{Transform} & \textbf{Coefficient Augmentation} & \textbf{\shortstack[c]{Types of \\ Misspecification Error}} \\
\midrule

Translation & Fourier & Group real and imaginary parts & Negligible error due to approximate exchangeability. \\
\midrule

\multirow{3}{*}{\shortstack[c]{Spatial \\ Permutation}}
    & Haar & Group by orientation per layer & \multirow{3}{*}{\shortstack[c]{Likelihood 
    constrained \\ misspecification  due to collaging \\ effect resolved by data.}} \\
    & Gabor & \multirow{2}{*}{\shortstack[c]{Group by filter}} & \\
    & Learned  & \\
\midrule

Rotation (90°) & Haar & Group horizontal and vertical details & Negligible error due to approximate exchangeability. 
\\
\midrule

\multirow{3}{*}{\shortstack[c]{Rotation \\ (Arbitrary)}}
    & Fourier & Group by frequency components & \multirow{3}{*}{\shortstack[c]{Arbitrary rotations are \\ Irrelevant Artifacts.}} \\
    & Gabor & Group over orientation per filter & \\

\midrule
Scaling & Fourier & Group similar frequency components & Negligible error due to approximate exchangeability. \\


\midrule
\multirow{2}{*}{Symmetrization} & Haar & \multirow{2}{*}{\shortstack[c]{Append negated \\ coefficients}}   & \multirow{2}{*}{\shortstack[c]{Recoverable Artifacts \\ after inspection}} \\
    & Gabor  & \\
\bottomrule
\end{tabular}
\label{tab: image_data_augment_mapping}
\end{table}




\subsection{Testing}
\label{section:testing}

\subsubsection{Independence}
\label{section:independence assumption}

The prior model assumes all of the coefficients are mutually independent  (see Section \ref{section:prior model}). To investigate this assumption, 
we estimate the pairwise covariances between each of the coefficient groups. We first create a bootstrap estimate of the sample covariance matrix, $C$. We analyze the relative magnitude of the off-diagonal entries by calculating the relative Frobenius norm = $\frac{||C - \text{Diag}(C)||_F}{{\text{Trace}(C)}}$. We also consider the cosine distance between the principal components of $C$ and the nearest unit vectors. 
The closer the principal components are to the unit vectors, the closer the covariance matrix is to the diagonal, so the smaller the cosine distances. Small cosine distances suggest there is a slight change of representation that decorrelates the coefficients since multiplication by the matrix of principal components diagonalizes $C$.

\subsubsection{Hypothesis Testing}

Statistical hypothesis tests determine if there is enough evidence to reject a hypothesis ($H_0$). Evidence against the hypothesis is measured by a $p$-value, the probability, under $H_0$, of observing a sampled test statistic as, or more, extreme than the observed statistic. The $p$-value is compared to a pre-determined threshold ($\alpha$) to decide whether the observed result would be implausibly extreme under the hypothesis. 
If the $p$-value is above the threshold, then there is not enough evidence to reject $H_0$. Large sample sizes reduce variance, so, for a fixed test statistic, the $p$-value decreases in sample size.

In this paper, we test the hypotheses:

\begin{enumerate}
\vspace{-0.0 in}
\item[] \textbf{$H_0$}: the empirical data is drawn from a generalized-gamma scale mixture of normals with parameters $r, \eta, \vartheta$.
\vspace{-0.05 in}
\item[] \textbf{$H_1$}: the empirical data is drawn from a different distribution.
\vspace{-0.0 in}
\end{enumerate}

A coefficient group ``fails" the hypothesis test if it is implausibly extreme under $H_0$ for any choice of parameters. Else, the coefficient group ``passes". Groups that pass are statistically indistinguishable, under our chosen test statistic, from samples generated under $H_0$. 

\subsubsection{Kolmogorov-Smirnov Testing}

We use a one-sample Kolmogorov–Smirnov (KS) test to evaluate whether a sample, with elements assumed to be i.i.d, is drawn from a member of the family. The test statistic:
\begin{equation}
    D_n = \sup_{x} \left| F_{n}(x) - F(x) \right|
    \label{eqn:KS Stat}
\end{equation}
is based on the cumulative distribution (CDF) of the null distribution, $F(x)$, and the empirical CDF of the sample, $F_{n}(x)$. Smaller values of the test statistic mean the empirical CDF is closer to the true CDF, so the KS statistic measures the discrepancy between the distributions. It evaluates the quality of a fit.

The KS test is nonparametric and does not make any assumptions about the underlying distribution of the sample. The distribution of the test statistic under the null, the Kolmogorov Distribution, is only dependent on the number of samples and not on the null distribution. There is a power-law relationship between the value of the KS statistic that produces a $p$-value of 0.05 and the sample size. Doubling the sample size decreases the cutoff by a factor of 4. We use the exact Kolmogorov distribution in our analysis.  

We need to specify the hypothetical CDF, $F$, to find the KS statistic. Since computing the closed-form of $F$ is intractable, we integrate numerically. We developed a one-dimensional quadrature procedure that rearranges the order of integration to exploit the closed form for the generalized gamma CDF. We use a cubic spline to approximate the CDF between quadrature approximations. See Appendix \ref{app:CDF_quadrature} for further details.

Finding the exact KS statistic for large coefficient groups proved computationally intensive. Instead, we approximated the empirical CDF by sorting the samples and taking $10^5$ regularly spaced values. Our experiments showed that the error in the KS statistic due to this approximation was negligible, with the precision to 4 decimal places.

\subsubsection{Interpretation of the KS statistic}

As introduced in Section \ref{section:data treatment}, we categorize misspecification errors by their downstream effects on inferential decision making. This framing motivates two uses of the KS statistic: first, to determine whether there is sufficient evidence to reject the prior, and second, to evaluate the degree of misfit in the prior. The statistical interpretation uses thresholds that vary as a function of sample size to determine whether observed discrepancies between model and data could be plausibly explained by randomness. When working with many samples, the plausibility standard is often very stringent. 
Sometimes the KS statistic may be implausibly large, even if the same fits could be practically useful since the inferential errors they produce are small. This occurs when the computed CDF is close enough to the true CDF to ensure that all misspecification errors are \textbf{negligible}. 

To check whether a member of the prior model could explain the empirical data, we conduct multiple hypothesis tests for various values of $r, \eta, \vartheta$. We do not apply any multiple hypothesis testing corrections. Multiple testing corrections are not needed for two reasons. First, we are not working with a fixed alternative, so there is no well-defined notion of family-wise error rate or false discovery proportion to control. Second, we do not expect that the actual data-generating process for the samples can be described by any member of this family, so we expect to always reject $H_0$ given enough data. Converting from a discrepancy to a $p$-value determines whether there is enough evidence to reject the model, not whether the model is plausible. Cases that pass the test should be considered statistically indistinguishable from samples generated by the model, not declared as evidence for the model. From this perspective, converting from a statistic to a test simply provides an alternate interpretation of the degree of discrepancy observed. We will use this framing.

\subsection{Fitting}
\label{section:fitting}

\subsubsection{Coarse Search Procedure}
We adopt a grid search procedure over the $r, \eta$ space, to characterize the regions, rather than points, of best fit.
We compute and store the CDF for each value of $r, \eta$ on an initial grid spanning $0\leq r \leq 20$ and $-1.5\leq \eta \leq 20$. 
This region covers the plausible parameters and is efficient. Finer spacing supports more precise estimates of best-fit parameters. We adopt a finer spacing of 0.1 for $0\leq r \leq 10$ and $-1.5\leq \eta \leq 10$. For $r>10$ and $\eta > 0$, we increase the spacing to 1.

\subsubsection{Scale}
\label{section:scale}
The complete set of hyperparameters includes the scale parameter $\vartheta$. Figures \ref{fig:ks kl plot} and \ref{fig:scale similarity} show that, for a fixed scale, the shape parameters are well-resolved by the prior variance, but are difficult to identify without a fixed scale. Testing with a constant scale $\vartheta = 1$ showed that $r$, $\eta$ pairs that matched the variance had much lower KS values. The variance of the prior can be computed analytically via iterated expectation, and is related to the $\vartheta$ by:
\begin{equation}
   \text{Var}[\pi(r, \eta, \vartheta = k)] = k\text{Var}[\pi(r, \eta, \vartheta = 1)]
   \label{eqn:Scale Variance Equation}
\end{equation}
 See Appendix \ref{app:nth_moment} for more details. 
 
 Using Equation \eqref{eqn:Scale Variance Equation}, the scale parameter $\vartheta$ could be chosen for each ($r, \eta$) pair, so that the variance of the proposed distribution matches the empirical variance observed in the sample. Sometimes, the empirical variance was dominated by a few outliers. To avoid overfitting to outliers, the full shape parameter grid was extended to include nine different $\vartheta$ values. Searching over scale directly is difficult since the range of possible values varies tremendously in the datasets considered. Instead, we consider the sample variance computed after excluding $t$ extreme values from each of the tails. We search over $t$, with choices of $[0, 25, 50, 75, 100, 150, 200, 250, 300, 350, 500]$. The value of $t$ that augments the sample variance (and in turn, $\vartheta$) such that the KS-test is minimized is denoted by $t_0$.

\subsubsection{Refined Grid Search}
After finding the initial best fit ($r_0, \eta_0, \vartheta_0$) from the initial grid search, we performed a refined search on a smaller grid of points centered around the initial best fit point. In regions with small $r$ $(< 0.02)$ or small $\eta$ $(\eta <0)$, a finer spacing is used to account for the increased sensitivity. We refine our estimate $\vartheta_0$ by performing a search over $t$, centered at $t_0$. The new values of $\vartheta$ arise from matching the sample's variance after excluding $t^*$ tails, with $t^*=\max(t_0+\Delta t, 0)$. Here, $\Delta t$ is found by searching over $[-100, -75, -50, -25, 0, 25, 50, 75, 100]$. This produces our best fit parameters ($r^*, \eta^*, \vartheta^*)$.


\section{Results}
\label{section:results}

\subsection{Do we fit?}

\begin{figure}
    \begin{center}
     \centering
     \includegraphics[width=.85\linewidth]{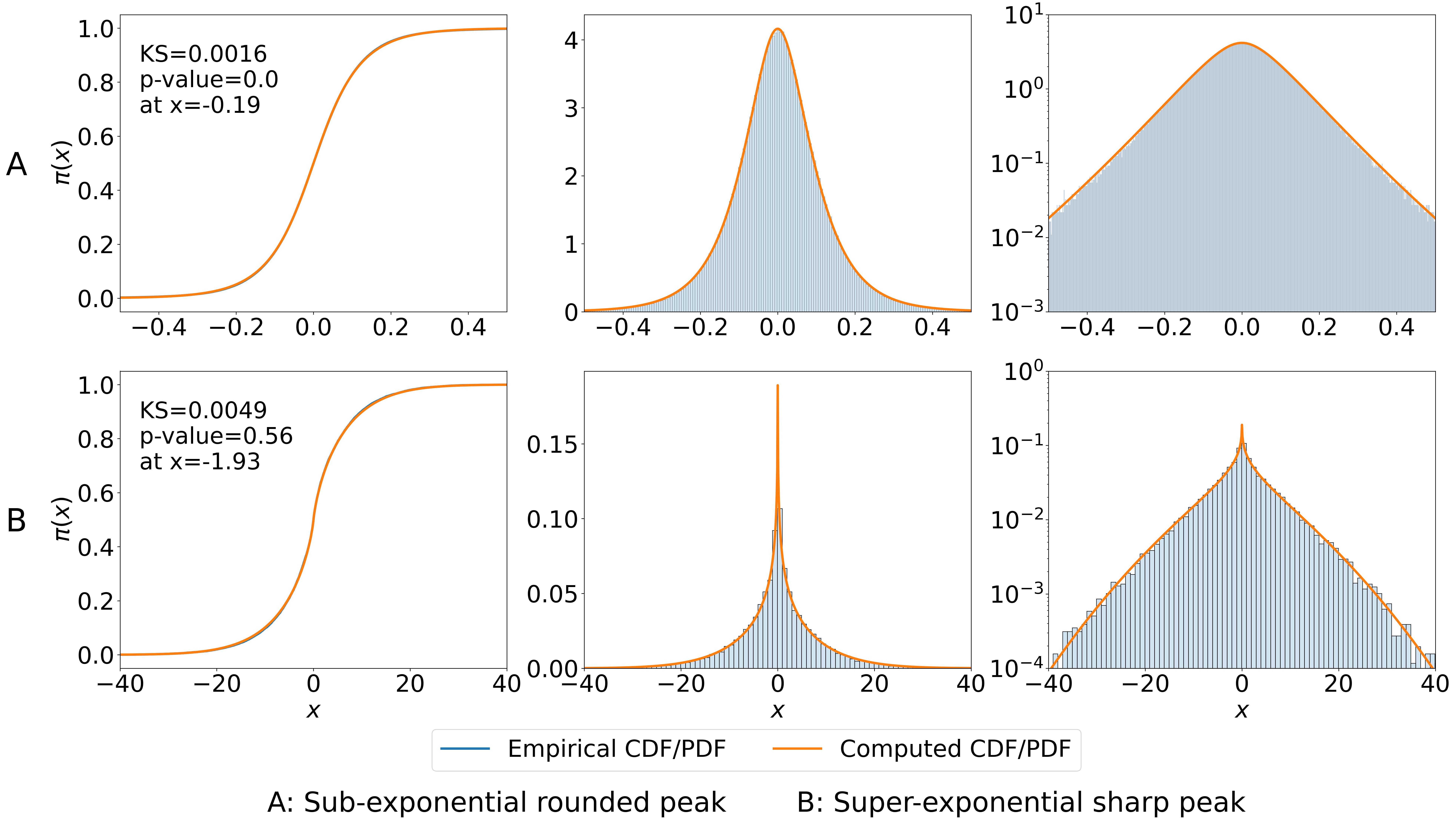}
  \end{center}
  \caption{Two high-quality fits to distributions with different peak and tail behavior. The top row (pastis, Fourier, gray, Band 29) illustrates that the prior can model distributions with a rounded peak and sub-exponential tails ($r=0.1, \eta=12, \vartheta=5.72e-24$). The log plot shows the tails are not Gaussian. The second row (coco indoor, Haar wavelet, diagonal, red, layer 4) shows a distribution with an extremely sharp and non-differentiable peak but with super-exponential tails ($r=20, \eta=-1, \vartheta= 231$). Neither distribution can be modeled well by the other commonly used priors.}
\label{fig:diverse_fits}
\end{figure}

The prior model achieves high-quality fits for the majority of datasets. Figure \ref{fig:diverse_fits} shows that the hierarchical model can fit distributions that do not fit a standard model (rounded peak with sub-exponential tails or non-differentiable peak with super-exponential tails). The hierarchical prior consistently outperforms the other commonly used priors that it includes as special cases (Gaussian, Laplace, Student's $t$) since it allows separate control of tail and peak behavior. In total, 85.0\% of the empirical distributions that fit the hierarchical prior could not be fit by any of the classical models. This indicates that separate control of peak and tail behavior is needed to model these distributions.  

We manually categorize the fits for each coefficient block by their KS statistic (Quantitative Criteria), $p$-value (Statistical Criteria), and visual observation of the CDF, PDF, and $\log(\text{PDF})$ fits (Qualitative Criteria). We assign each coefficient group to one of five categories: statistically pass, practically pass, borderline, interesting failure, or trivial failure. \textit{Statistically pass} indicates that the discrepancies observed between the empirical CDF and the hypothesized model can be plausibly accounted for by randomness. Then, the empirical and computed best-fit distributions are statistically indistinguishable. In cases that \textit{statistically fail}, the errors encountered may still be tolerably small. This occurs in the large sample size limit, when the statistical test is very stringent, so it is highly sensitive to small discrepancies that would produce \textbf{negligible} misspecification errors. To capture this nuance, we make the distinction between \textit{statistically pass} ($\alpha=0.05$) and \textit{practically pass}.
Table \ref{tab: fit categorization} contains a more detailed description of the categories. See Appendix \ref{app:fit_categorization} for example fits belonging to each category.

\begin{table}
\caption{Fit categorization criteria based on KS statistic and visual fit quality.}
\centering
\small
\begin{tabular}{@{}p{3.0cm}p{4.2cm}p{7.5cm}@{}}
\toprule
\textbf{Category} & \textbf{Quantitative Criteria} & \textbf{Qualitative Criteria / Notes} \\ \midrule
Statistically Pass &
$p$-value $> 0.05$, more than 100 samples &
Discrepancies, measured by the KS statistic, could be attributed to sampling error, and the visual fit is good. \\[2pt]
Practical Pass &
KS statistic small ($< 0.01$, sometimes up to $0.015$) &
Peak and tail behavior are captured well. None of the failure modes apply; though statistically significant deviations exist, misspecification is negligible. \\[2pt]
Borderline &
KS statistic $< 0.02$ &
Very close to a good fit but has a small failure mode; may still be usable. Failure mode matches one of the types listed below. \\[2pt]
Interesting Failure &
--- &
Unclear if the distribution can be modeled within the family; often involves a pronounced peak. \\[2pt]
Trivial Failure &
--- &
Feature clearly outside the prior family; subtypes include asymmetry, multimodality, double inflection point, and spike-and-slab behavior. \\ \bottomrule
\end{tabular}
\label{tab: fit categorization}
\end{table}

\begin{figure}[h]
    \begin{center}
    \includegraphics[width=\linewidth]{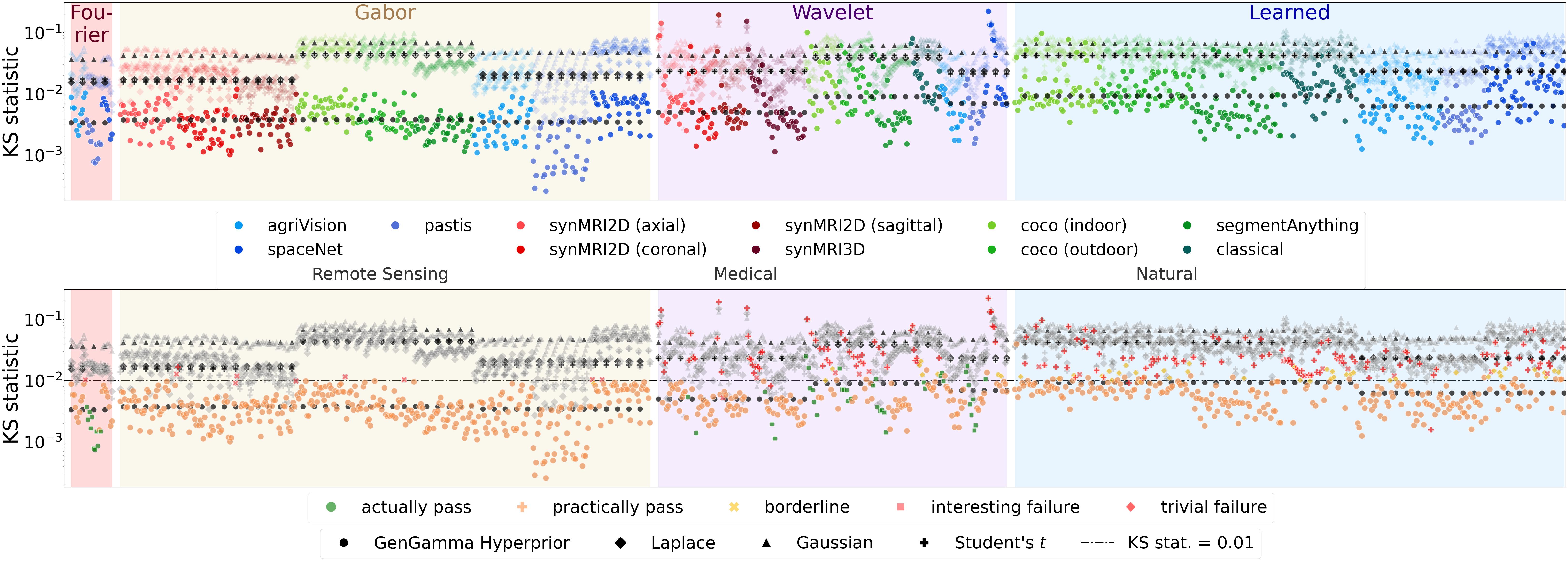}
    \end{center}
  \caption{KS statistics organized by transform and dataset type, colored by dataset. The horizontal position of each point within the dataset indicates the frequency of the Fourier and Haar wavelet transforms. The marker types represent the different priors. The black markers indicate the median KS statistic within that dataset type for the specified transform. }
\label{fig:aggregate ksstat}
\end{figure}

Table \ref{tab: results} and Figure \ref{fig:aggregate ksstat} indicate that the prior model produces high-quality fits. Of the 1071 fits, 71.8 \% pass with 15.5\% classified as a statistical pass and 56.3\% as a practical pass. 
Table \ref{tab: results} indicates that the median KS statistic is less than 0.01 for nearly every dataset-transform combination. 
Our prior model performs better than standard priors for 91.5\% of the 1071 coefficient blocks. Of the remaining 8.5\%, 87.9\% comprise trivial failures and cannot be modeled well by any of the priors. 
The first panel of Figure \ref{fig:aggregate ksstat} highlights that the median KS statistic of our prior is approximately an order of magnitude better than other priors across all dataset types.
This behavior is expected relative to the Gaussian and Laplace, as these distributions comprise special cases of the prior model included in the search space. 

The median samples column of Table \ref{tab: results} indicates the stringency of the statistical pass standard. We find that, in cases where we pass, the KS statistics are small enough such that the tests would make the same conclusions for samples of size $10^4-10^6$.

\begin{table}
\caption{Fit quality across dataset types and transforms alongside median samples. The combined pass percentage is the sum of the fits classified as either \textit{statistical} or \textit{practical} passes.}
\small
\centering
\begin{tabular}{l l c |c c|c}
\toprule
\textbf{Transform} & \textbf{Dataset} & \textbf{\shortstack{Median\\Samples}} & \textbf{\shortstack{Median\\KS Stat.}} & \textbf{\shortstack{Stat. Pass\\\(\alpha=0.05\) (\%)}} & \textbf{\shortstack{Combined Pass\\(\%)}} \\
\midrule
\multirow{4}{*}{\textbf{Fourier}} 
  & agriVision & 2.6e6 & 0.005 & 2.3 & 95.5 \\
  & pastis & 2.6e5 & 0.002 & 90.9 & 100.0 \\
  & spaceNet & 1.8e6 & 0.005 & 0.0 & 100.0 \\
  \cmidrule(lr){2-6}
  \rowcolor{gray!20} \whitecell & Remote Sensing & 5.5e5 & 0.003 & 34.2 & 98.4 \\
\midrule
\multirow{10}{*}{\textbf{Gabor}} 
  & synMRI2D (axial) & 1.8e8 & 0.005 & 0.0 & 95.2 \\
  & synMRI2D (coronal) & 1.4e8 & 0.003 & 0.0 & 95.2 \\
  & synMRI2D (sagittal) & 9.7e7 & 0.004 & 0.0 & 97.6 \\
  \cmidrule(lr){2-6}
  \rowcolor{gray!20} \whitecell & Medical & 1.4e8 & 0.004 & 0.0 & 96.0 \\
  \cmidrule(lr){2-6}
  & coco (indoor) & 6.6e7 & 0.006 & 0.0 & 97.0 \\
  & coco (outdoor) & 6.6e7 & 0.004 & 0.0 & 95.8 \\
  & segmentAnything & 2.6e8 & 0.003 & 0.0 & 100.0 \\
  & standardTesting & 2.4e6 & 0.008 & 0.0 & 57.1 \\
  \cmidrule(lr){2-6}
  \rowcolor{gray!20} \whitecell & Natural & 6.6e7 & 0.004 & 0.0 & 94.5 \\
  \cmidrule(lr){2-6}
  & agriVision & 2.6e8 & 0.003 & 0.0 & 100.0 \\
  & pastis & 1.6e7 & 0.001 & 0.6 & 100.0 \\
  & spaceNet & 1.6e8 & 0.007 & 0.0 & 92.9 \\
  \cmidrule(lr){2-6}
  \rowcolor{gray!20} \whitecell & Remote Sensing & 1.6e8 & 0.003 & 0.2 & 97.6 \\
\midrule
\multirow{14}{*}{\textbf{Haar}} 
  & synMRI2D (axial) & 4.2e6 & 0.007 & 0.0 & 76.2 \\
  & synMRI2D (coronal) & 3.7e6 & 0.004 & 9.5 & 80.9 \\
  & synMRI2D (sagittal) & 2.4e6 & 0.005 & 0.0 & 81.0 \\
  & synMRI3D & 1.6e6 & 0.005 & 7.1 & 69.0 \\
  \cmidrule(lr){2-6}
  \rowcolor{gray!20} \whitecell & Medical & 2.9e6 & 0.005 & 4.8 & 75.3 \\
  \cmidrule(lr){2-6}
  & coco (indoor) & 2.6e5 & 0.013 & 25.0 & 50.0 \\
  & coco (outdoor) & 3.9e5 & 0.011 & 25.0 & 53.1 \\
  & segmentAnything & 1.8e6 & 0.004 & 25.0 & 85.2 \\
  & standardTesting & 9.2e3 & 0.013 & 9.5 & 47.6 \\
  \cmidrule(lr){2-6}
  \rowcolor{gray!20} \whitecell & Natural & 4.1e5 & 0.008 & 24.0 & 62.6 \\
  \cmidrule(lr){2-6}
  & agriVision & 1.7e6 & 0.005 & 15.3 & 100.0 \\
  & pastis & 1.5e5 & 0.004 & 57.1 & 100.0 \\
  & spaceNet & 1.6e6 & 0.012 & 4.7 & 53.1 \\
  \cmidrule(lr){2-6}
  \rowcolor{gray!20} \whitecell & Remote Sensing & 7.0e5 & 0.007 & 24.0 & 84.4 \\
\midrule
\multirow{8}{*}{\textbf{Learned}} 
  & coco (indoor) & 2.0e8 & 0.009 & 0.0 & 64.5 \\
  & coco (outdoor) & 2.0e8 & 0.010 & 0.0 & 57.8 \\
  & segmentAnything & 7.9e8 & 0.005 & 0.0 & 81.2 \\
  & standardTesting & 7.1e6 & 0.014 & 0.0 & 24.1 \\
  \cmidrule(lr){2-6}
  \rowcolor{gray!20} \whitecell & Natural & 2.0e8 & 0.009 & 0.0 & 58.2 \\
  \cmidrule(lr){2-6}
  & agriVision & 7.9e8 & 0.005 & 0.0 & 73.3 \\
  & pastis & 4.9e7 & 0.004 & 0.0 & 93.8 \\
  & spaceNet & 4.8e8 & 0.013 & 0.0 & 42.9 \\
  \cmidrule(lr){2-6}
  \rowcolor{gray!20} \whitecell & Remote Sensing & 4.8e8 & 0.006 & 0.0 & 66.2 \\
\bottomrule
\rowcolor{gray!20} \textbf{\large Overall}
  &  &  6.5e7 & 0.005 & 7.3 & 83.5 \\
\bottomrule
\end{tabular}
\label{tab: results}
\end{table}

\subsection{For what Dataset and Transforms?}

For different dataset types and transforms, we select fits representative of the average, best, and worst cases observed. We examine how fits vary as a function of group number. For the Haar wavelet and Fourier transforms, we can order the filters by frequency, and the groups correspond to the layers and bands, respectively. 

\begin{figure}
    \begin{center}
        \begin{minipage}{1\textwidth}
            \centering
            \hspace*{0.7cm}
            \includegraphics[width=.6\linewidth]{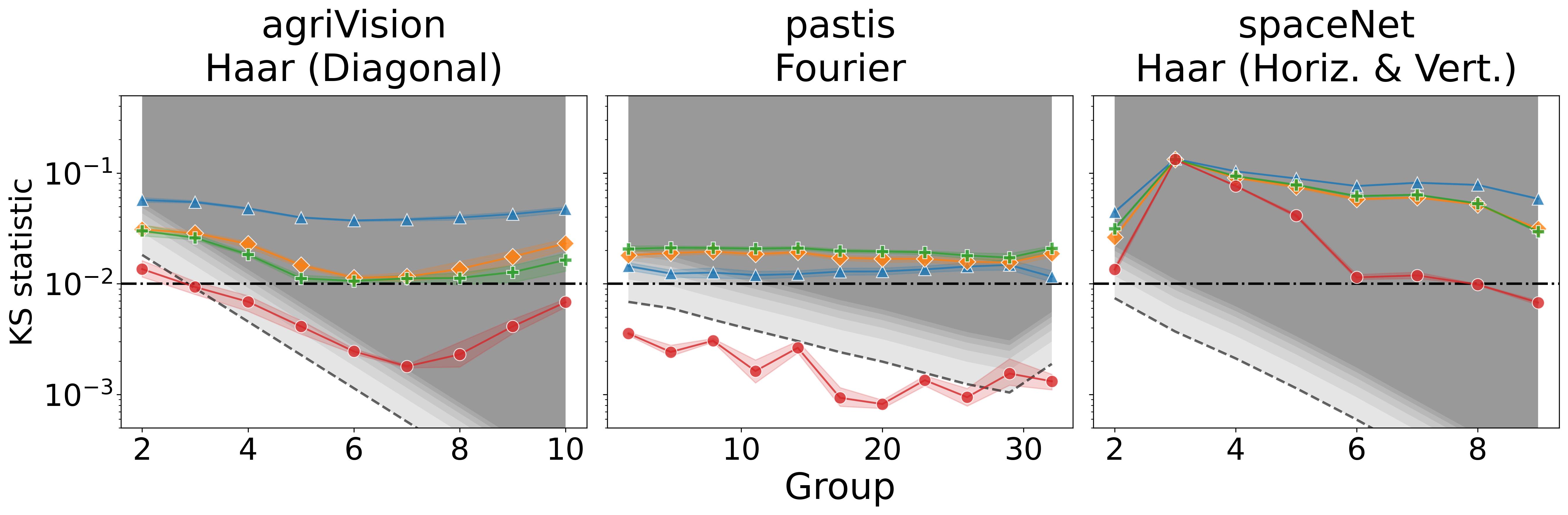}
            \includegraphics[width=.75\linewidth]{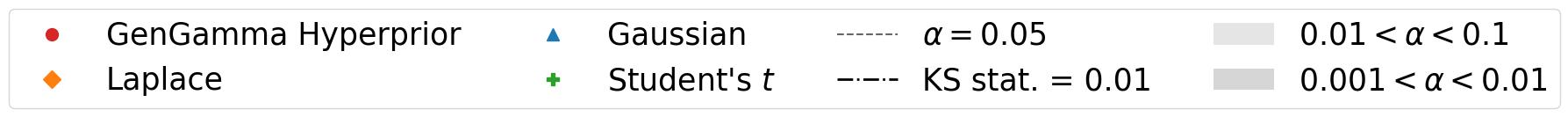}
        \end{minipage}
    \end{center}
    \caption{KS statistics for select dataset-transform combinations with the remote sensing datasets (\textit{agriVision}, \textit{pastis}, \textit{spaceNet}), aggregating over channel (Red, Green, Blue).}
    \label{fig:remote ksstat}
\end{figure}

Figure \ref{fig:remote ksstat} shows three examples from the \textbf{remote sensing image datasets}. In the left plot (\textit{agriVision}), we see the typical trend. The KS statistics first decrease as the CDFs become more resolved, and then slightly increase at higher layers.  
In the middle plot (\textit{pastis}), the KS statistics nearly all lie below the statistical pass threshold. After applying a Fourier transform, the \textit{agriVision} and \textit{spaceNet} trends are similar to \textit{pastis}, with the difference being that more of them are classified as \textit{practical} passes.
Finally, the third plot shows some fits that constituted trivial failures. We attribute this to outlier images within the \textit{spaceNet} dataset that include large bodies of water, which produce regions in the images where the color intensities are very close to constant. As a result, a Haar wavelet that evaluates averaged differences on a neighborhood fully contained in a body of water returns values very close to zero. The larger the neighborhood, the more pixel values averaged, so the closer the difference will be to zero. If the neighborhood is too large, then it may not fit entirely within the body of water. This could explain the observed jump in KS statistics between group numbers 2 and 3. This gives rise to spike-and-slab distributions that cannot be modeled by generalized gamma scale mixtures of normals. 
Together, this explains the varied breakdown of statistical and practical pass percentages for \textit{pastis} ($57.1\%$ + $42.9\%$), \textit{agriVision} ($15.3\% + 84.7\%$), and \textit{spaceNet} ($4.7\% + 48.4\%$) after applying the Haar wavelet transform.
In contrast, applying the Fourier to all three datasets produced high-quality fits with a combined pass percentage exceeding $95\%$ across all three datasets.

\begin{figure}[h]
    \begin{center}
        \begin{minipage}{1\textwidth}
            \centering
            \hspace*{0.7cm}
            \includegraphics[width=.6\linewidth]{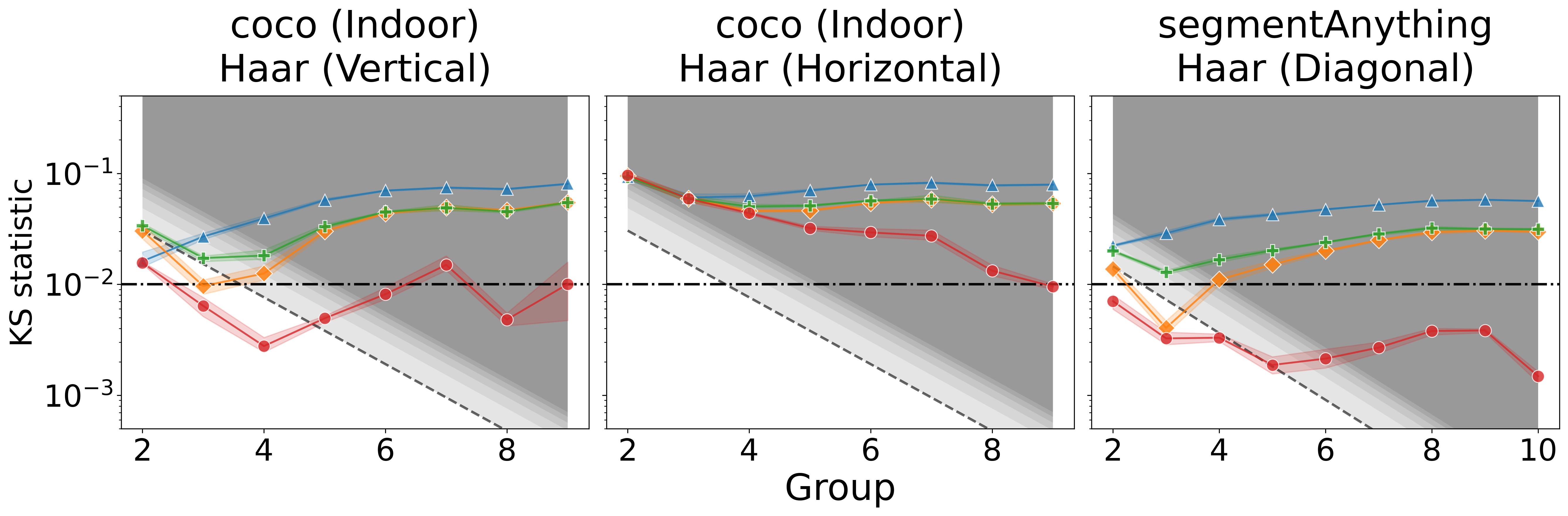}
            \includegraphics[width=.75\linewidth]{figure/results/ksstat_legend.jpg}
        \end{minipage}
    \end{center}
    \caption{KS statistics for select dataset-transform combinations with the natural image datasets (\textit{coco}, \textit{segmentAnything}), aggregating over channel (Red, Green, Blue).}
    \label{fig:natural ksstat}
\end{figure}

Figure \ref{fig:natural ksstat} shows three examples from the \textbf{natural image datasets}. The orientation of the Haar wavelet transform has a significant impact on the quality of the fit. 
In all three datasets, the vertical and diagonal orientations produced better fits than the horizontal orientations, particularly in the lower layers. We speculate this is because natural images tend to be aligned with the horizon, and the horizontal details are obtained using vertical differencing (see Figure \ref{fig:filters_and_wavelet_layers}). The horizontal coefficients have the largest variance and smallest kurtosis of all three orientations. 
There was no discernible difference between \textit{coco (outdoor)} \textit{coco (indoor)}, so we only show the KS statistics for \textit{coco (indoor)}.

\begin{figure}[h]
    \begin{center}
        \begin{minipage}{1\textwidth}
            \centering
            \hspace*{0.7cm}
            \includegraphics[width=.7\linewidth]{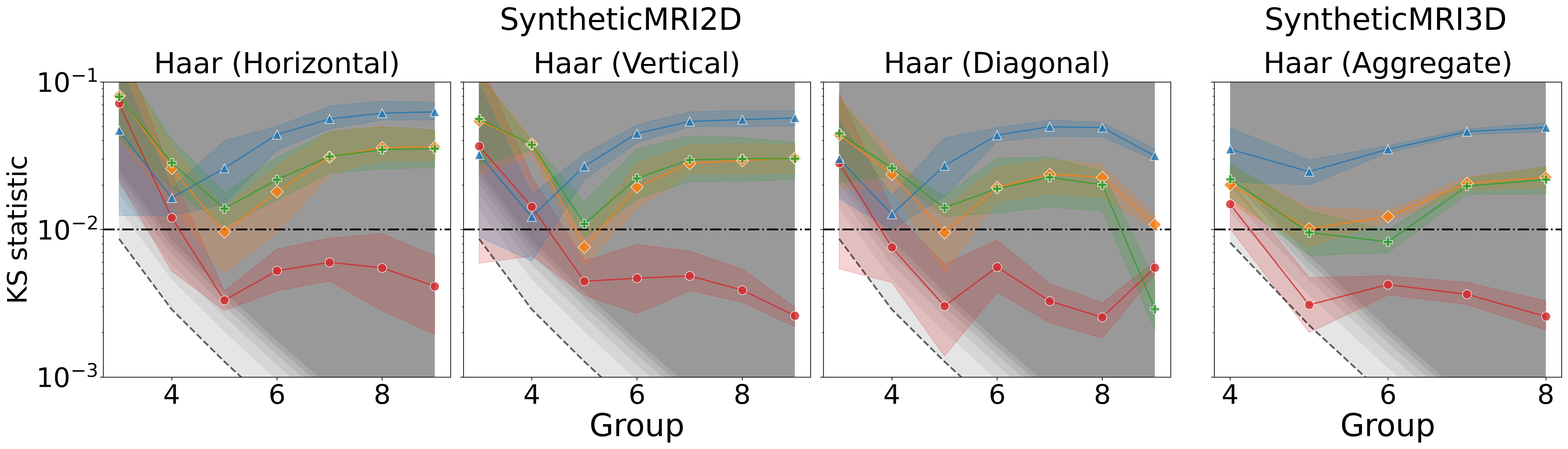}
            \includegraphics[width=.75\linewidth]{figure/results/ksstat_legend.jpg}
        \end{minipage}
    \end{center}
    \caption{KS statistics for select dataset-transform combinations with the medical image datasets (\textit{syntheticMRI2D}, \textit{syntheticMRI3D}), aggregating over brain cross-section (2D) and Haar wavelet orientation (3D).}
    \label{fig:medical ksstat}
\end{figure}

Figure \ref{fig:medical ksstat} shows all the KS statistics from the \textbf{medical image datasets}. Unlike the previous plots, which aggregate over color channels, here, we only work with grayscale images and aggregate over the different cross-sections of the brain. For \textit{syntheticMRI2D}, this refers to the axial, coronal, and sagittal slices of the brain. There is much greater variability in the KS statistics across the orientation of the cross-section than observed across color in the previous examples.
For \textit{syntheticMRI3D}, the Haar wavelet transform produces seven different ``orientations", depending on the order in which we perform differencing and averaging operations. The KS statistic decreases as the layer number (i.e.~frequency of the wavelet) increases.

\begin{figure}
    \begin{center}
        \begin{minipage}{1\textwidth}
            \centering
            \hspace*{0.7cm}
            \includegraphics[width=.9\linewidth]{figure/results/ksstats_gabor_boxplot_by_params.jpg}
        \end{minipage}
    \end{center}
    \caption{KS statistics for the remote sensing and natural image datasets after applying the Gabor filters, aggregating over (wave number, aspect ratio) parameters. A single box plot represents the range of KS statistics for a particular dataset, with the colored points signifying the lower quartile KS statistic for the other priors (Laplace, Gaussian, Student's $t$).}
    \label{fig:gabor ksstat}
\end{figure}

For the \textbf{remote sensing} and \textbf{natural} image datasets, we apply the Gabor filters to produce the KS statistics in Figure \ref{fig:gabor ksstat}. Observing frequency to be the most influential parameter on KS statistics, we group by the other two filter parameters (wave number, aspect ratio). Gabor filters produced many more coefficients than the Fourier or Haar wavelet transforms, with median sample sizes in the $10^7-10^8$ range. These large sample sizes enforce more stringent passing standards, so we observe few statistical passes even though the best-fit KS statistics remain small. We observed median KS statistics below 0.01 for all dataset and parameter groups, and particularly tight fits (median KS statistics near 0.001) for the \textit{pastis} dataset.

\begin{figure}[h]
    \begin{center}
        \begin{minipage}{1\textwidth}
            \centering
            \hspace*{0.7cm}
            \includegraphics[width=.9\linewidth]{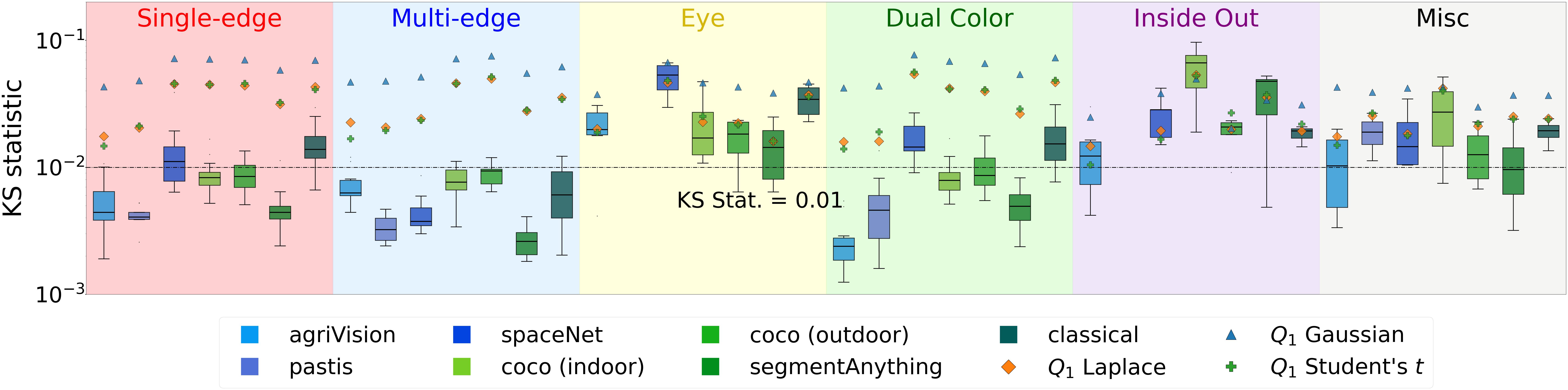}
        \end{minipage}
    \end{center}
    \caption{KS statistics for remote sensing and natural image datasets after applying learned filters, aggregating over filter group. A single box plot represents the range of KS statistics for a dataset, with the colored points signifying the lower quartile KS statistic for other priors (Laplace, Gaussian, Student's $t$).}
    \label{fig:learned ksstat}
\end{figure}

We also apply the learned filters to the \textbf{remote sensing} and \textbf{natural} images to produce the KS statistics in Figure \ref{fig:learned ksstat}. The learned filters do not admit a well-defined ordering, so we aggregate by the filter groups illustrated in Figure \ref{fig:filters_and_wavelet_layers}. If a particular box plot is missing (for example, \textit{pastis}-eye), that means that no filters in the corresponding filter group passed the skew test. Like with Gabor, the large sample size resulted in no statistical passes despite low KS statistics in the single and multi-edge cases. The single-edge and multi-edge filters, which are most similar to the Haar wavelet, produce the best fits with nearly all median KS statistics below $0.01$. Outside of those two filter groups, there is more variability in the quality of the fit. For these filters, we observe more multimodal and asymmetric coefficient distributions. 


\subsection{Where do we fit?}

\begin{figure}[h]
    \centering
    \includegraphics[width=\linewidth]{figure/results/where_parameters_lie.jpg}
    \hfill

    \caption{Best fit parameters categorized by fit category (see Table \ref{tab: fit categorization} for criteria) and by transform. The same points are represented in both $r, \eta$ and $r, \beta$ space, with the $\ell_pER$ in red (along $\eta=0$), $r=1$ line in blue, and $\eta=-1$ line in purple as specified in Figure \ref{fig:parameter space map}. The two plots in the third column display the Regions of best fit in $r$, $\beta$ space. The top plot shows the pastis dataset under the Fourier transform, and the bottom shows the spaceNet dataset under a Haar wavelet transform. The colors in these plots represent the group number, a proxy for the frequency of the filter, with darker regions representing lower groups (lower frequency). Only the regions of best fit in cases that statistically or practically pass are shown. Stars in the rightmost column represent the best fit $r,\eta$ pair when allowing scale to vary.}
    \label{fig:aggregate point and region examples}
\end{figure}

The first two columns of Figure \ref{fig:aggregate point and region examples} show the best-fit ($r, \eta$) and ($r, \beta)$ pairs. Recall that $\beta = (\eta+1.5)/r$. Because $\beta$ is positive, it is used in the following plots to allow log spacing of the axis. For a fixed $r$, increasing $\beta$ increases $\eta$, so a larger $\beta$ leads to a rounder peak. Most of the Haar and Gabor points are underneath the $\eta = 0$ line with varying $r$, meaning the peak of the distribution is not differentiable, and the density may diverge at 0. The majority of the learned filters have small $r$, typically less than $0.3$, with varying $\eta$ values, meaning the tail decay for these distributions is sub-exponential. Finally, the Fourier points are distributed evenly between the four quadrants, and there are no clear trends due to the lower number of tests. Notably, almost all of the best fit points lie above $\eta=0$, with those below lying close to center of all the quadrants. These indicate that the observed distributions are far from Gaussian, since Gaussian distributions correspond to large positive $r$ or large positive $\eta$. 

The plot in the second column at the top indicates that the quality of the best fit point is related to its location. The points that are classified as failures (both interesting and trivial) tend to lie on the border of the search grid or below the $\eta = -1$ line. Because the prior distribution has infinite density at $x=0$ when $\eta \leq -1$, the numerical CDF is highly sensitive to small parameter changes in this region. Similarly, the large or sometimes infinite variances of the low $r$ cases $(r<0.1)$ also make numerical calculation challenging. In these regions, the instability in computing the CDF can lead to worse fits. The passing cases (both statistically and practically) are distributed across the interior of the remaining search region, indicating that we usually find good quality fits where we can reliably compute CDFs.

The point plots in the first two columns of Figure \ref{fig:aggregate point and region examples} do not provide enough evidence to make sharper claims about the best-fit regions as they relate to changes in coefficient group, representation, or dataset. Point estimates are highly sensitive to the specific KS statistics and tell an incomplete story. 
In practice, there is a neighborhood of parameters that produces similar fits. We define the set of $r, \eta$ that either produce KS statistics that lie below the 0.05 p-value threshold, or lie below our practical pass KS statistic threshold of 0.01 to be the \textit{region of best fit}. The third column of Figure \ref{fig:aggregate point and region examples} shows two representative examples. These regions appear convex in $r$,$\frac{1}{\beta}$ space, with approximately linear boundaries, so they can be represented in $\log{r}$, $\log{\beta}$ space after a simple transformation. These regions summarize the range of acceptable parameter choices for each coefficient group. 

The breadth of these regions is explained by the near unidentifiability of $r,\eta$ under exchanges of the scale parameter $\vartheta$. As discussed in Section \ref{section:prior model}, changes in $\vartheta$ can account for changes in $r$, $\eta$ without making large changes to the distribution (see Figures \ref{fig:scale similarity} and \ref{fig:fit kurtosis variance}). The region plots show a projected version of the 3D region in $r$, $\beta$ space for the best-fit distributions. If these ranges intersect the $\ell_p$-Equivalent Region ($\ell_pER$), then an $\ell_p$ style prior could be adopted instead of the generalized-gamma scale mixture of normals. The two examples shown in the rightmost column of Figure  \ref{fig:aggregate point and region examples} highlight the wide variation in identifiability observed. In the first, $r$ and $\eta$ are poorly identified. In the second, they are precisely specified.





\subsection{Independence}
\label{section:independence results}

Table \ref{tab: independence results} shows the results from the independence tests described in Section \ref{section:independence assumption}. The results are calculated individually for each dataset-transform combination, and the average for each group is displayed. The relative Frobenius norms are fairly small ($\approx$ 1e-5) for all of the dataset types. This means that, relative to the variance of the components, the magnitude of the pairwise covariances is fairly low. Within each transform, the dataset types performed similarly, so the biggest insights are gained from comparing transforms. The median cosine distances for the Haar wavelet and Fourier components are also low, which indicates that the covariance matrices are approximately diagonal. Together, these tests do not detect evidence against the independence of the components after applying the Fourier and Haar wavelet transforms. Conversely, the learned filters produce coefficients with higher cosine distances, so they detect more evidence of dependence. The Gabor filters produce metrics between the Fourier transforms and learned filters. This indicates that there is some dependence, but not as much as was seen with the learned filters. The 90$^{th}$ percentile and the maximum cosine distances after applying the Fourier transform indicate that the covariance is not exactly diagonal. These larger dependencies appear in higher-order principal components that explain less of the total variance. Most of the principal components are reasonably close to the unit vectors. The $90^{\text{th}}$ percentile and the max. $\cos$ distance after applying the Haar wavelet transform are lower than the Fourier transform, but share the same interpretation.

\begin{table}
\caption{Summary of structural similarity metrics (relative Frobenius norm of group covariance differences; median, tail and maximum cosine distances) computed for each (Transform, Dataset, Channel) combination and averaged over the dataset type.}
\centering
\small
\begin{tabular}{l l c | c | c c c}
\toprule
\textbf{Transform} & \textbf{Dataset Type} & \textbf{\shortstack{Num.\\Groups}} &
\textbf{\shortstack{Relative\\Frob. Norm}} &
\shortstack{\\Median}&
\shortstack{\textbf{Cos Dist.}\\90th Perc.} &
\shortstack{\\Max} \\
\midrule
\textbf{Fourier} & Remote Sensing & 29 & 2.9e-05 & 0.0310 & 0.1281 & 0.2416 \\
\cmidrule(lr){2-7}
\rowcolor{gray!20} \whitecell
                & \whitecell Aggregate & \whitecell 29 & \whitecell 2.9e-05 & \whitecell 0.0310 & \whitecell 0.1281 & \whitecell 0.2416 \\
\midrule
\textbf{Gabor}   & Medical & 42 & 3.0e-05 & 0.0818 & 0.2821 & 0.5855 \\
                & Natural & 42 & 3.0e-05 & 0.0204 & 0.2765 & 0.5942 \\
                & Classical & 42 & 3.0e-05 & 0.0842 & 0.3219 & 0.5871 \\
                & Remote Sensing & 42 & 3.0e-05 & 0.1097 & 0.2961 & 0.5823 \\
\cmidrule(lr){2-7}
\rowcolor{gray!20} \whitecell
                & \whitecell Aggregate & \whitecell 42 & \whitecell 3.0e-05 & \whitecell 0.1524 & \whitecell 0.323 & \whitecell 0.601 \\
\midrule
\textbf{Haar} & Medical & 6.56 & 2.4e-05 & 0.0059 & 0.0284 & 0.0299 \\
                & Natural & 8.33 & 1.3e-05 & 0.0028 & 0.0312 & 0.0354 \\
                & Classical & 7 & 1.5e-05 & 0.0036 & 0.0125 & 0.0149 \\
                & Remote Sensing & 8 & 1.7e-05 & 0.0045 & 0.0157 & 0.0188 \\
\cmidrule(lr){2-7}
\rowcolor{gray!20} \whitecell
                & \whitecell Aggregate & \whitecell 7.82 & \whitecell 1.6e-05 & \whitecell 0.004 & \whitecell 0.0252 & \whitecell 0.0285 \\
\midrule
\textbf{Learned} & Natural & 62.33 & 3.2e-05 & 0.2265 & 0.4196 & 0.8876 \\
                & Classical & 54 & 3.2e-05 & 0.2220 & 0.4207 & 0.9557 \\
                & Remote Sensing & 49.33 & 3.2e-05 & 0.2129 & 0.5157 & 0.8919 \\
\cmidrule(lr){2-7}
\rowcolor{gray!20} \whitecell
                & \whitecell Aggregate & \whitecell 55.57 & \whitecell 3.2e-05 & \whitecell 0.22 & \whitecell 0.4609 & \whitecell 0.8991 \\
\bottomrule
\end{tabular}
\label{tab: independence results}
\end{table}

\section{Discussion}
\label{section:discussion}

Our primary contribution is the answer to the question: Is the prior realistic? We find compelling examples where the prior model defined in Section \ref{section:prior model} produces high-quality fits to empirical data spanning varying sources,  kinds and representations. 
These examples suggest that, for properly specified parameters, the generalized gamma scale mixtures of normals could be used to produce priors with properly calibrated marginal distributions and could be used for realistic uncertainty quantification.

The fact that wavelet-like filters (Haar, Gabor, learned single-edge, learned multi-edge) performed well across all tested datasets suggest that wavelet-like transforms may produce coefficient distributions near our prior family in a wide variety of imaging contexts. 
Even the natural image datasets, with greater variability of objects, perform better with wavelet-like filters.
In contrast, other filters may produce non-standard distributions that cannot be modeled well by symmetric, unimodal priors.

The quality of fit depends on parameters being well-specified. The regions of plausible parameters are large and vary depending on factors including the frequency of the filter and the sample size. In $40.9\%$ of cases, for the correct choice of scale parameter, we find the prior is equivalent to $\ell_p$ priors.
Since we do not identify any systematic trends that hold across all dataset types or transforms, we recommend future work adopt an Empirical Bayes' approach, or fit to an actual dataset, to specify the parameters of the prior. This overlap with the $\ell_pER$ validates some approaches within the computational methods literature. However, our findings also suggest that focusing on this region alone is myopic.


\subsection{Limitations}

\subsubsection*{Choice of Test Statistic}

The testing pipeline is built around the KS statistic. The KS statistic, which measures the maximum \textit{absolute} deviation between CDFs, is more sensitive to the peak of the distributions than their tails. Distributions with dramatically different rates of tail decay may share a small KS statistic, while inducing very different regularizers. Using other metrics that better capture the tail behavior, like the kurtosis or the Anderson-Darling statistic, could prove useful. Finding distributions that pass the intersection of two hypothesis tests could provide more precise fits for the parameters by separately constraining the fit to the bulk and to the tails of the distribution. 

\subsubsection*{Role of the scale parameter $\vartheta$}
\begin{figure}
    \begin{center}
     \centering
     \includegraphics[width=.9\linewidth]{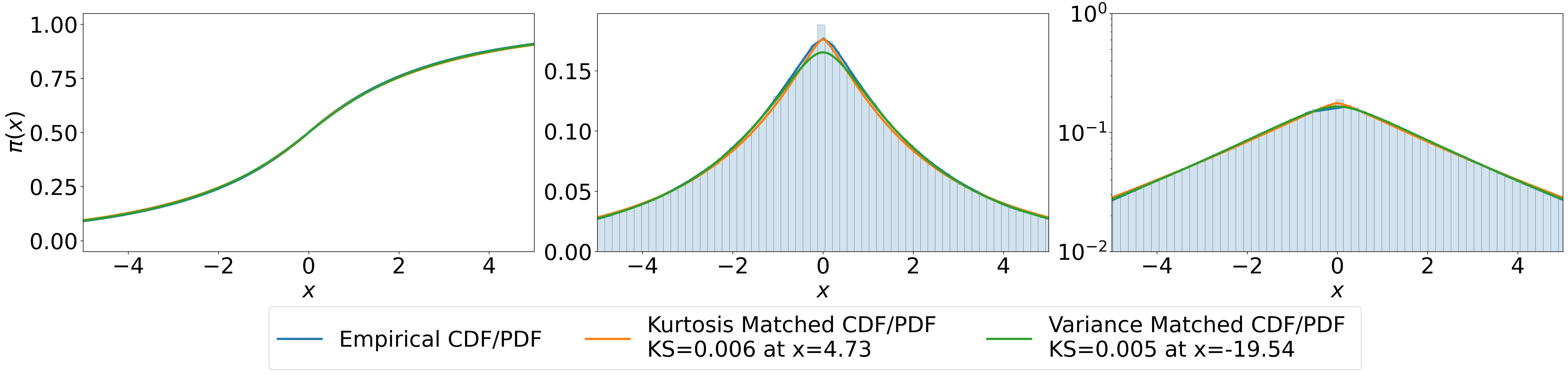}
  \end{center}
  \caption{Two similar fits to a sample (agriVision, Wavelet Diagonal, Layer 6) after picking the scale parameter by matching to the variance ($r=0.1, \eta=3.8, \vartheta=5.99e-17$) and to the kurtosis ($r=0.4, \eta=-0.1, \vartheta=5.96e-01$).  Note that the vastly different values of the parameters produce similar KS statistics. Matching the variance before the grid search better captures peak behavior, so the max. deviation occurs in the tails. Matching the kurtosis modeling the tails better, with the max. deviation occurring closer to the peak.}
\label{fig:fit kurtosis variance}
\end{figure}

As described in Section \ref{section:scale}, we search over a subset of potential scale parameters as determined by the empirical variance. This limits the set of parameters that could plausibly fit the empirical distribution to a set of manifolds in $r, \eta, \vartheta$ space. The error in estimating the empirical variance due to outliers propagates through the scale parameter. Moreover, there are other reasonable ways to pick the scale parameter, for example, by matching to the empirical kurtosis instead of the variance. This could produce similar fits, as assessed by the KS statistic, with vastly different parameters (see Figure \ref{fig:fit kurtosis variance}). Different applications may require different methods of specifying the scale parameter.

\subsubsection*{Limited validation of the independence assumptions} 
Our model assumes both that the coefficients within a block are independent and that the blocks of coefficients are independent of each other. The lack of pairwise correlations between blocks is not sufficient to prove independence. 
Additionally, we do not test for within-block independence. Because we grouped coefficients representing similar features, the within-block independence assumption is less likely. As discussed in Section \ref{section:representation}, for each of the transforms, coefficients representing similar features are less likely to be independent.

\subsection{Next Steps}

These methodological limitations do not change our core conclusion: \textit{the prior model} fits real data. Rather, they indicate that it may be possible to get even better fits by choosing a different test statistic and more precisely fitting the scale parameter.  

Our work does not explain \textit{why} certain coefficient blocks fit better than others. Factors could include the following properties of images:

\begin{enumerate}
    \item Variability of subject, with natural images being most variable. 
    \item Plausible invariances, with remote sensing images exhibiting the most symmetries.  
    \item Presence of outliers, driven by single images. For example, the \textit{spaceNet} dataset contains aerial shots of large bodies of water that introduce a spike-and-slab distribution with high-frequency filters. 
    \item Effect of subsampling the sorted data, which distorts its variance and kurtosis.
\end{enumerate}

\vspace{0.05in}

Future work should:

\begin{enumerate}
    \item Test the generality of our results by exploring data from different domains (e.g.~audio) or modalities (e.g., EEG, CT), and different representations, 
    like learned filters from other contemporary ML models. 
    \item Improve the testing procedure by incorporating multiple test statistics, or by changing the test statistic used to produce more precisely specified best-fit parameters. Reasonable alternatives include the Anderson-Darling statistic, or the model likelihood. We did not adopt the likelihood in this study since its sampling distribution is unknown. Even if the range of plausible parameters remains large, improving the stability of best-fit parameters could allow investigations into the relationship between coefficient type and typical distribution. For instance, how do the likelihood-maximizing parameters vary as the frequency of the filter varies?
    \item Address the stability of posterior inference to model misspecification, both within and outside the family of priors, since it is likely that most applied studies that adopt this prior family will not fully resolve the parameters. 
    \item Test the independence assumption more comprehensively, both within-block in addition to across blocks.
\end{enumerate}

\section*{Conflicts of Interest} The authors have no conflicts of interest to declare.

\section*{Acknowledgments} The authors thank Daniela Calvetti and Erkki Somersalo for their inspiration and support over many years working on generalized gamma scale mixtures. They also thank Ani Adhikari for her encouragement and guidance in recruiting student collaborators.

\bibliographystyle{siam}
\bibliography{references.bib}

\newpage
\appendix
\section{Survey of Computational Methods Literature}
\label{app:survey}

Here, we list the 18 papers used in our survey of the computational methods literature. Specifically, we examine the types of examples used to validate the methods proposed. We study the types of examples used to validate the method. Computed or synthetic examples denote cases where the signal or image was artificially generated. Classic examples refer to standard images like the Shepp-Logan Phantom or the MIT Cameraman image, which have an element of realism but are not necessarily representative of real data. The \textit{Motivation} column indicates the primary reason the numerical method was proposed -- either to rectify computational gaps in the literature (e.~g.~generalizing IAS methods), or with specific applications in mind (e.~g.~cerebral source localization).

\begin{table}[H]
    \centering
    \caption{Survey of papers using a conditionally Gaussian prior model with (generalized) gamma hyperpriors.}
    \label{tab:validation_survey}
    \renewcommand{\arraystretch}{1.25}
    \begin{tabular}{p{10cm}cp{2cm}c}
        \toprule
        \textbf{Title} & \textbf{Year} & \textbf{Examples} & \textbf{Motivation} \\
        \midrule
        Hypermodels in the Bayesian imaging framework \cite{calvetti2008hypermodels} & 2008 & Synthetic & Computational \\
        Conditionally Gaussian hypermodels for cerebral source localization \cite{calvetti2009conditionally} & 2009 & Synthetic & Application \\
        Sparse Bayesian image Restoration \cite{babacan2010algorithm} & 2010 & Classic & Application \\
        A hierarchical Krylov–Bayes iterative inverse solver for MEG with physiological preconditioning \cite{calvetti2015hierarchical} & 2015 & Synthetic & Application \\
        Bayes meets Krylov: Statistically inspired preconditioners for CGLS \cite{calvetti2018bayes} & 2018 & Synthetic & Computational \\
        \rowcolor{gray!20} Brain activity mapping from MEG data via a hierarchical Bayesian algorithm with automatic depth weighting \cite{calvetti2019brain} & 2019 & Real and Synthetic & Application \\
        Hierarchical Bayesian models and sparsity: $\ell^2$-magic \cite{calvetti2019hierachical} & 2019 & Synthetic & Computational \\
        Sparse reconstructions from few noisy data: analysis of hierarchical Bayesian models with generalized gamma hyperpriors \cite{calvetti2020sparse} & 2020 & Synthetic & Computational \\
        Sparsity promoting hybrid solvers for hierarchical Bayesian inverse problems \cite{calvetti2020sparsity} & 2020 & Synthetic & Computational \\
        Overcomplete representation in a hierarchical Bayesian framework \cite{pragliola2020overcomplete} & 2020 & Synthetic & Computational \\
        A variational inference approach to inverse problems with gamma hyperpriors \cite{agrawal2022variational} & 2022 & Synthetic & Computational \\
        Hierarchical ensemble Kalman methods with sparsity‑promoting generalized Gamma hyperpriors \cite{kim2022hierarchical} & 2022 & Synthetic & Computational \\
        Hierarchical Ensemble Kalman Methods with Sparsity-Promoting Generalized Gamma Hyperpriors \cite{kim2022hierarchical} & 2022 & Synthetic & Computational \\
        Sparsity promoting reconstructions via hierarchical prior models in diffuse optical tomography \cite{manninen2024dot} & 2023 & Synthetic & Computational \\
        \rowcolor{gray!20} Sequential image recovery using joint hierarchical Bayesian learning \cite{xiao2023algorithm} & 2023 & Real and Synthetic  & Computational \\
        Leveraging joint sparsity in hierarchical Bayesian learning \cite{glaubitz2024leveraging} & 2024 & Synthetic & Computational \\
        Efficient sampling for sparse Bayesian learning using hierarchical prior normalization \cite{glaubitz2025efficient} & 2025 & Synthetic & Computational \\
        Efficient sparsity‑promoting MAP estimation for Bayesian linear inverse problems \cite{lindbloom2025efficient} & 2025 & Classic & Computational \\
        \bottomrule
    \end{tabular}
\end{table}

\newpage
\section{Gabor and Learned Filter Bank}
\label{app:learned_gabor_filters}

\begin{figure}[!htb]
    \begin{center}
     \centering
     \includegraphics[width=.8\linewidth]{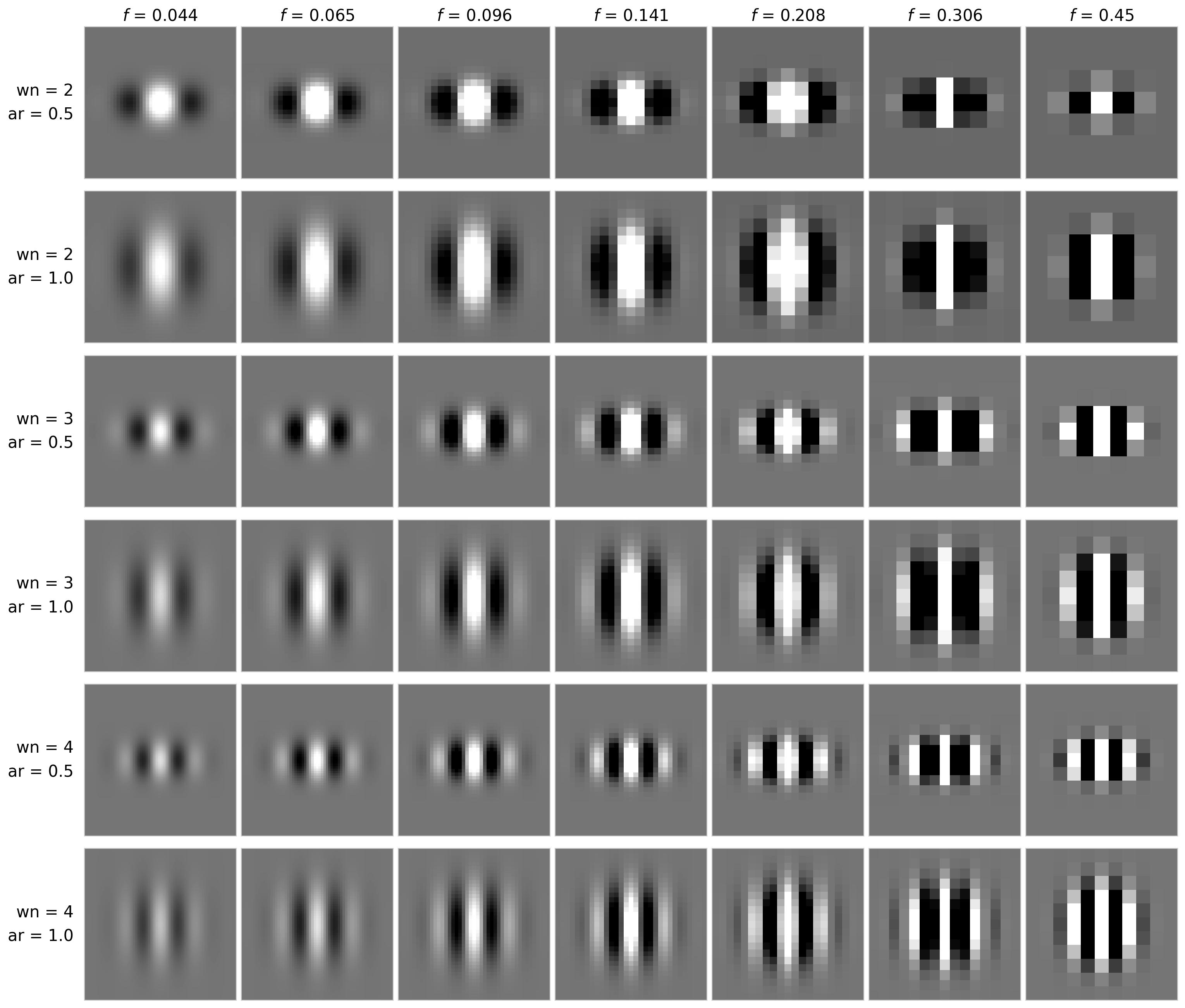}
  \end{center}
  \caption{Visualization of all 42 Gabor filters used in the vertical orientation, organized by frequency (f), wave number (wn), and aspect ratio (ar).}
  \label{fig:gabor_filters_all}
\end{figure}

\begin{figure}
    \begin{center}
     \centering
     \includegraphics[width=.8\linewidth]{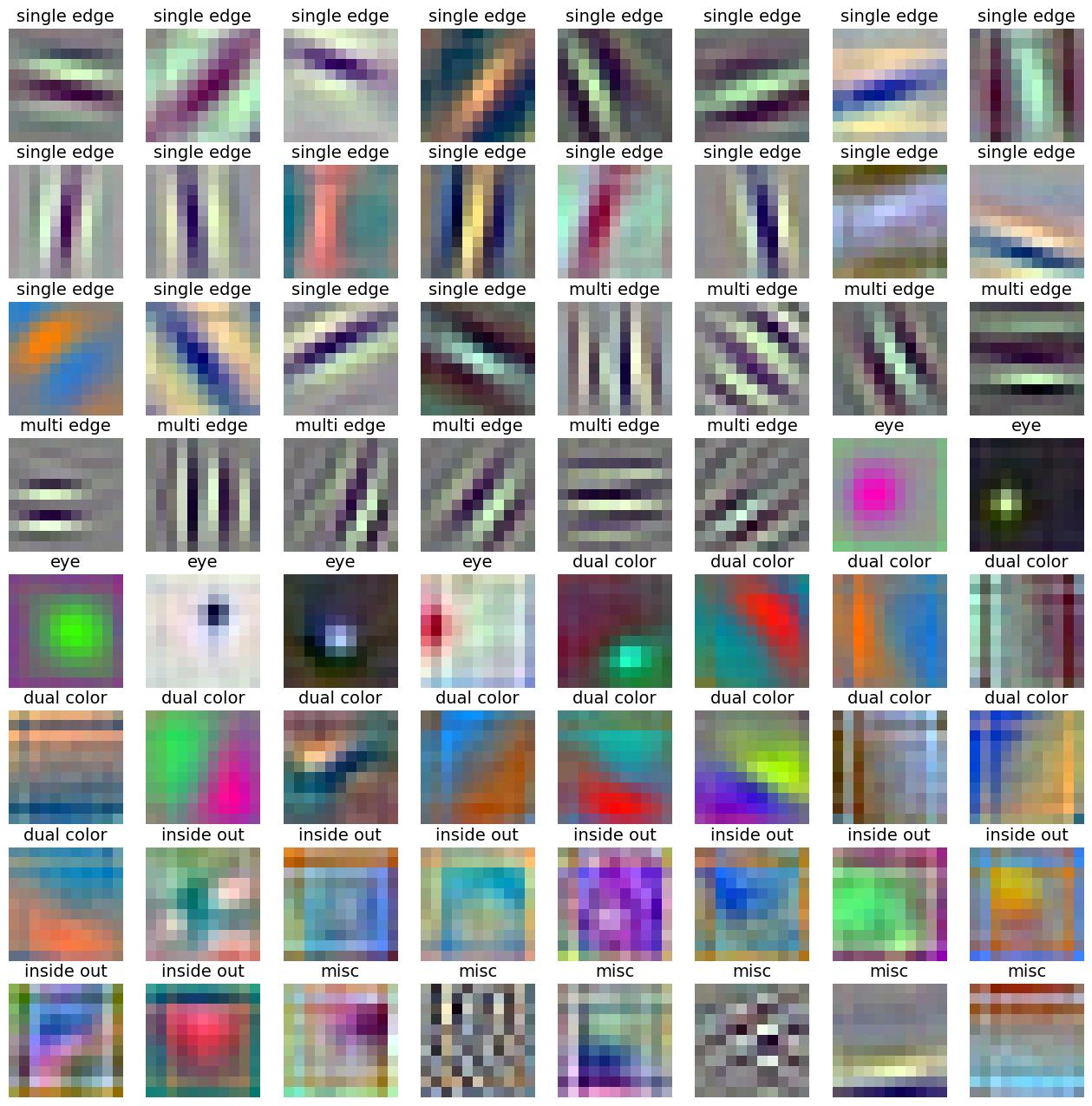}
  \end{center}
  \caption{Visualization of all 64 learned filters used, normalized 0-1, from the first layer of AlexNet with their filter category.}
  \label{fig:learned_filters_all}
\end{figure}

\newpage

\section{Misspecification Error Types}
\label{app:misspec_error_types}
\subsubsection*{Negligible} The error in the prior is small. 
    
In this case, the modeled prior is close to some notion of a gold-standard prior. As long as the inferential pipeline is continuous in the divergence used to measure errors in distribution, then sufficiently small misspecification errors in the prior will induce acceptably small errors in the recovered images. These errors generally correspond to assumptions that simplify modeling by electing to ignore detailed aspects of the gold standard. 
This is an acontextual error standard since it does not consider the downstream influences of prior errors.

\subsubsection*{Likelihood Constrained} The errors in the prior do not induce large errors in inference.
    
Posterior inferences are informed by both the prior and the likelihood. In practice, priors are usually introduced to regularize inference. Here, it is worth considering the form of our prior model in more detail, since the regularizing effect of the prior depends on its structure and interaction with the likelihood. 

Ideally, a prior should regularize inference by exploiting all available prior information. In practice, such priors are complex and can only be constructed with highly expressive models. There has been success in using generative modeling to learn high-dimensional joint distributions with complex structures. These types of models excel at resolving common image features \cite{bhadra2020medicalimagereconstructionimageadaptive}. 
However, highly expressive distribution fits are often underdispersed, so they may discount actual irregularities when applied. In a medical application, an abnormal structure might be rejected by the prior, leading to hallucinated or spurious reconstructions. If the prior constrains recovery too closely to common cases, then the associated inferential errors are difficult to distinguish post-hoc. More generally, it is difficult to evaluate the calibration of highly expressive, high-dimensional models, so it is difficult to anticipate and account for the degree and kind of inferential errors they induce. 

In contrast, we focus on a single-tiered hierarchical family that uses only three parameters per block of coefficients. Models of this kind are highly constrained, so are necessarily misspecified. Our model family assumes both independence among the coefficients and a parametric form for the marginals of each coefficient. We do not expect actual independence between all pairs of coefficients, nor do we have a generative reason to justify our parametric form. Nevertheless, assumptions of this kind are commonly employed in inverse imaging \cite{dias2003parametricindependence, Figueiredo2003parametricindependence, sanyal2025parametricindependence}. They are justified by separating the role of the likelihood and the prior. Since most forward models are ill-conditioned, some form of regularization is needed; however, regularization is usually not needed along all directions at once. Instead, typical likelihoods tightly constrain inference along some directions, while remaining essentially uninformative along others. 

In this setting, simplified parametric priors aim only to regularize inference along the subset of directions that are poorly resolved by the likelihood, while relying on the likelihood for reconstruction of most substantive image content. Often, prior assumptions, like sparsity in a wavelet domain, aim to regularize textural aspects of the image that, when combined with the likelihood, constrain the space of plausible inferences enough to eliminate excessive variation generated by amplification of measurement noise. These priors do not target complete realism. Instead, they should be sufficiently overdispersed to leave substantive recovery to the likelihood, while specific enough to eliminate uncertainty along directions the likelihood cannot stably resolve. Maintaining this separation of roles is desirable since it distinguishes aspects of inference that are data-informed from aspects that are assumption-constrained. Relaxing the prior along directions resolved by the likelihood increases posterior uncertainty, but largely in tails that are ignored by posterior summaries. So, we are willing to admit errors in the prior that induce substantial overdispersion, provided those errors would be reliably resolved by conditioning on the observed data. 

\subsubsection*{Irrelevant Artifacts} The errors in the prior induce large errors in inference; however, these errors do not influence decisions made in response to inference.

Errors in the prior that are not corrected by the likelihood may be passed on to the posterior summaries used for inference. These errors are inconsequential if discarded by the summary. More broadly still, the decisions drawn from inferential summaries may be invariant to errors that do not change the user's interpretation. For example, if the user is primarily interested in image classification, then their final decisions may be invariant to some large image transformations. These include registration errors, translation, some spatial permutations, and some magnifications. In fact, modern classifiers often train on augmented datasets that randomly transform the input image ensemble so that the classifier learns to ignore large, but irrelevant, transformations. In Section \ref{section:grouping_via_induced_exchangeability}, we will show that some data treatment choices (e.g.,~assigning coefficient groups) can be formally exchanged with specific data augmentations, so may be justified if the associated augmentations would leave downstream decision making unchanged.

\subsubsection*{Resolvable Artifacts} The errors in the prior induce large errors in inference; these errors would hinder decision making but can be corrected by post-processing.

Errors of this kind can be identified by the downstream user as artifacts. Some artifacts may be corrected by applying a more informative prior post-hoc. For example, most users could classify color-inverted or vertically inverted images by hand. The actual priors needed to resolve these artifacts are often subtle, and contextual, since they rely on global image cues that are not easily discerned at a pixel or coefficient level, so require carefully tailored joint distributions. However, such a prior could be trained separately and applied post-hoc to correct such artifacts. We admit these artifacts on the grounds that a user could apply a two-stage inference procedure that aims to first resolve the image up to correctable artifacts, then to correct the artifacts. In Section \ref{app:grouping via induced exchangeability}, we will show that some data treatment choices can be formally exchanged with specific data augmentations that could be easily inverted by a second-stage prior. 

\newpage

\section{Transform Overview}
\subsubsection*{Fourier} The Fourier transform converts functions on the real spatial/pixel domain to functions on the complex frequency domain. Each image is recovered from its transform via a linear combination of waves, where the coefficients of the combination equal the transform of the original image. Each wave is specified by a frequency vector that orients the wave and fixes its wavelength. 
Each coefficient corresponds to a wave, so it is both determined by, and affects, the entire image. The magnitude and complex phase of the coefficient determine the amplitude and phase of the wave. 

The Fourier transform is a common choice in signal processing since it is mathematically fundamental and computationally efficient. Quasi-sparsity is achieved when the image can be reconstructed using a few key wave functions. Images of this kind may exhibit strongly periodic behavior over large regions.
For example, satellite images of farmed fields exhibit large patches of periodic behavior because fields are often planted in long, regularly spaced rows. The independence assumption under the Fourier transform means that the amplitude and phase of the waves with different wavelengths and orientations are independent. 

\subsubsection*{Gabor} Gabor filters are defined by a sinusoidal carrier modulated by a Gaussian envelope, yielding filters that are localized in space while remaining selective to specific frequencies and orientations. Each Gabor coefficient reflects the presence and strength of a locally oriented oscillatory pattern, making these filters effective smooth edge detectors. We evaluate a bank of 42 Gabor filters with varying frequencies, aspect ratios, and wave numbers. Here, the wave number corresponds to the number of oscillations with non-negligible amplitude contained within the Gaussian envelope. We select aspect ratios of $0.5$ and wave numbers of $2$, $3$, and $4$ based on physiological observations of receptive fields in early visual cortex \cite{jones1987evaluation, usrey2003receptive}. An aspect ratio of $1$ is also included for conceptual contrast.

Gabor filters are well-suited to capturing oriented textures and mid-level features. This makes them particularly appealing for natural, remote-sensing, and medical images, which contain edges and structures at multiple scales. Quasi-sparsity is plausible when images contain only a limited number of prominent oriented structures per region, leading to mostly small coefficients punctuated by strong responses at edge locations. Independence under the Gabor representation implies that the presence and strength of distinct localized edge patterns are independent across space and orientation, an assumption that is reasonable for sufficiently separated features.

\subsubsection*{Haar}
Conversion to the Haar wavelet basis retains both spatial and frequency information while providing a discrete, multiscale representation of local image structure. Unlike the Fourier transform, each Haar coefficient is associated with a specific spatial neighborhood. Coefficients are highly interpretable, corresponding to spatially averaged differences evaluated in horizontal, vertical, or diagonal directions at different scales. The Haar transform decomposes an image into a sequence of layers, where earlier layers capture coarse, low-frequency structure and later layers encode finer details. Each layer beyond the first is subdivided into horizontal, vertical, and diagonal detail components. The first layer consists of a single coefficient equal to the average pixel intensity, which is fixed to zero by normalization. The Haar wavelet transform was applied to all of the datasets. Figure~\ref{fig:filters_and_wavelet_layers} illustrates the resulting layers and detail components for a test image.

Like Gabor filters, Haar wavelets are effective at representing local patterns, particularly edges. However, Haar wavelets capture these features using piecewise-constant basis functions rather than smooth oscillatory patterns. Hence, images that are approximately piecewise constant over large regions are quasi-sparse in the Haar wavelet basis, such as satellite imagery containing large bodies of still water. A localized transform is also essential in settings where certain regions of the image are uninformative or should be ignored, such as the black background present in MRI scans. Independence is plausible for Haar coefficients associated with spatially distant regions, but much less so for nearby or overlapping coefficients, such as adjacent edges at the same scale.

\subsubsection*{Learned} Finally, we tested the 64 convolution filters in the first layer of AlexNet.  Each filter was classified into one of six categories (see Figure \ref{fig:filters_and_wavelet_layers}). 
Because some of the filters produced highly idiosyncratic coefficient distributions, we ran a preliminary test to estimate the skew of the empirical distribution with 95\% confidence. For each dataset, we excluded the filters that produce distributions with confidence intervals that do not include 0. These cases clearly violate prior assumptions, such as unimodality or symmetry.

Unlike the Fourier and Haar wavelet basis functions, these filters are not chosen for their mathematical properties. Instead, their relevance is justified by their utility in image processing tasks. If the quasi-sparsity assumption holds, then each feature only appears in a few locations per image. Coefficients corresponding to distinct parts of the images, or to filters that detect very different features, are plausibly independent. 

\section{Grouping Arguments}
\label{app:grouping_args}
\subsection{Approximate Exchangeability}
\label{app:grouping via approximate exchangeability}
Recall that a block of coefficients is approximately exchangeable if all of the images generated by swapping the blocked coefficients are close to equally likely under the data-generating process. In this appendix, we discuss the coefficient groups that led to an approximately exchangeable group of coefficients. 

Rotating an image 90$^{\circ}$ corresponds to swaps in the horizontal and vertical details after applying the Haar wavelet transform. 
 If the probability of sampling an image and its 90$^{\circ}$ rotation are approximately the same, then the coefficients that swap roles under the rotation are approximately exchangeable. 
 This augmentation is justified for the remote sensing datasets. When rotated 90$^\circ$, the aerial view of a farmed field or a city still looks like a farmed field or a city.
Coefficients associated with diagonal details produced different marginals than horizontal or vertical details. Since the pixels are stored on a square lattice, there is no simple rotation to convert either the horizontal or vertical Haar wavelet components into the diagonal Haar wavelet components. 
We did not combine the horizontal and vertical components in the natural or medical images because the variables were not naturally exchangeable. Both medical and natural images admit features that are assigned a canonical orientation, and distinguish the horizontal from the vertical Haar wavelet components (e.g.,~the horizon line in an outdoor image).

Invariance to translation, or stationarity, is widely assumed in spatial statistics. Translating an image corresponds to grouping the real and imaginary parts of the Fourier transform coefficients. Swapping the real and imaginary components of a Fourier coefficient shifts the phase of the associated wave. When tested, the empirical marginal distributions of the real and imaginary parts of the complex coefficients were indistinguishable. This observation is not surprising for image sets that are not phase-registered, for instance, cropped patches from satellite imagery. 


\subsection{Induced Exchangeability}
\label{app:grouping via induced exchangeability}
For example, we group all coefficients produced by the same filter applied to different locations. By doing so, we assume that images related by a spatial permutation of their features are equally likely. This assumption is a stronger version of the translation symmetry assumption observed empirically in Section \ref{app:grouping via approximate exchangeability}. It corresponds to an augmentation procedure where the spatial labels assigned to coefficients are scrambled to produce a collage. 
 If the forward model maps collaged images to essentially disjoint regions of the data space, then, given data from the true image, the likelihood will still reliably constrain inference near the original, unaugmented image. 
In this case, we admit misspecification errors generated by collages as \textbf{likelihood constrained}. 
For the Haar wavelet, the corresponding coefficient augmentation combines all the coefficients from each detail type at each layer. For the learned filters, the corresponding coefficient augmentation groups all the coefficients of a particular filter. 

\begin{figure}[t!]
    \begin{center}
     \centering
     \includegraphics[width=.8\linewidth]{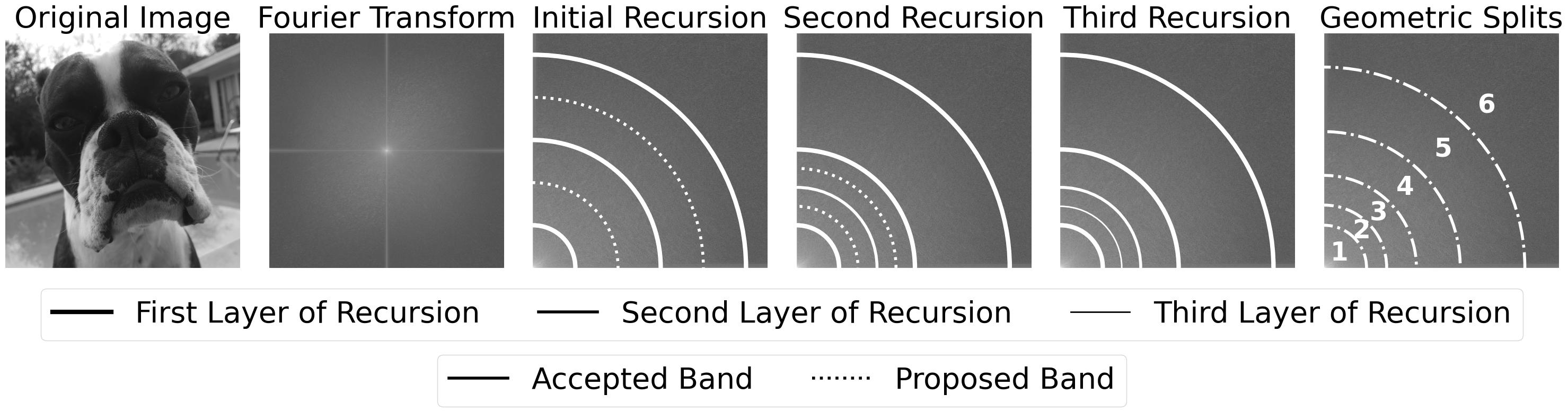}
  \end{center}
  \caption{Process for grouping coefficients into ``bands" after applying a Fourier transform, as described in \ref{section:grouping_via_induced_exchangeability}.}
\label{fig: fourier_bands_viz}
\end{figure}

We are also willing to augment some images with random rotations. This induces exchangeability between coefficients assigned to filters of the same type but with different orientations. Depending on the particular dataset, this augmentation can be justified as \textbf{negligible}, \textbf{irrelevant}, or \textbf{recoverable}. In the remote sensing examples, 90$^{\circ}$ rotations don't significantly alter the underlying image distribution, which produces negligible misspecification errors. This would discard orientation-specific information that is of little importance for most applications. For instance, a classifier designed to isolate objects from a satellite image should be able to identify most objects, no matter their cardinal orientation. So, even if given augmented data, the errors produced would, in most cases, be \textbf{irrelevant artifacts}. 

Random rotations combine Fourier coefficients with strictly the same wavelength, but different orientations. Both natural images and medical images admit canonical orientations. In these cases, large errors in orientation could be reliably inverted, so do not hinder downstream analysis. Small errors in orientation may be hard to recover, so may induce irresolvable artifacts. Depending on the downstream task, these may or may not produce relevant errors. For the reasons above, we chose to group over orientation for Gabor and Fourier, but did not apply Fourier to the medical and natural image datasets (the Fourier transform is unavailable in the medical setting irregardless, since the image must be cropped to exclude the background). In contrast to Gabor, Haar admits only 3 discrete orientations that cannot be interchanged without a large rotation, so we did not group over orientation for Haar except in Remote Sensing.
 
 The Fourier transform uses many unique wavelengths, so the Fourier coefficients must be further grouped into bands of similar wavelength. As shown in Figure \ref{fig: fourier_bands_viz}, the bands are created by partitioning the grid of frequency vectors into a sequence of wavelength intervals. We partition recursively. First, a wavelength interval is split into two halves. Then, we compute, the KS statistic between the groups of coefficients in each of the two new bands. If the statistic is small enough, the bands are deemed approximately exchangeable, and the process stops. If the KS statistic is large, each new band is passed through the recursive process. The interval boundaries are recorded, then fit to a geometric sequence. In all cases tested, the observed sequence of interval boundaries closely matched a geometric growth law. The fit is parameterized by a single multiplicative factor. The factor determines a regular partitioning in log wavelength. In practice, we do not attempt to fit long wavelength groups that include too few coefficients. 
 
 The corresponding image augmentation is spatial dilation by multiplicative factors close to one.  These augmentations transform the image by slightly zooming in or out. Slight zooms may be hard to recover. However, by continuity in distribution, and design of the partitioning procedure, we should introduce \textbf{negligible} misspecification errors. 

The prior is symmetric by construction. We can symmetrize the coefficient distributions by appending a negated copy to itself. The analogous image augmentation randomly inverts the intensity of each pixel. Given a linear forward model or change of basis, negating the output is the same as negating the input. During the exploratory data analysis (EDA) stage, if the distributions are slightly asymmetric, symmetrization is considered. It is only performed when we decide it would be obvious to invert, given a negated image (e.g. an MRI with a dark brain and a white boundary), and when the coefficient distributions produced by changing representation were close to, but not quite, symmetric. 
We only performed this augmentation on the medical imaging datasets.
 

\section{$n^{th}$ Moment Calculation}
\label{app:nth_moment}

\begin{equation}
    E[X^n] =
    \begin{cases}
        \dfrac{ (n-1)!! \,\vartheta^{\tfrac{n}{2}}
        \Gamma\!\left(\tfrac{\eta + 1.5 + \tfrac{n}{2}}{r}\right) }
        { \Gamma\!\left(\tfrac{\eta + 1.5}{r}\right)}, & n \text{ even}, \\[12pt]
        0, & n \text{ odd}.
    \end{cases}
    \label{eq:nth_moment_formula}
\end{equation}

For the following proof, recall that:
\[
    \pi(x|r,\eta,\vartheta) = \int_0^\infty \pi(x|\theta) \pi_{\text{hyper}}(\theta|r,\eta,\vartheta) d\theta
\]
where $\pi_{\text{hyper}}$ is a generalized gamma distribution. 

\begin{proof}
\begin{align*}
    E[X^n] 
        &= \int_x \int_\theta x^n \,\pi_{\mathcal{N}(0, \theta)}(x|\theta) 
            \,\pi_{\text{hyper}}(\theta|r,\eta,\vartheta)\, d\theta \, dx \\[6pt]
        &= \int_\theta \pi_{\text{hyper}}(\theta|r,\eta,\vartheta)
            \left( \int_x x^n \,\pi_{\mathcal{N}(0, \theta)}(x|\theta)\, dx \right) d\theta \\[6pt]
        &= \int_\theta \pi_{\text{hyper}}(\theta|r,\eta,\vartheta)\,
            E_{\mathcal{N}(0, \theta)}[X^n] \, d\theta
\end{align*}

The inner expectation is the $n^{th}$ moment of a Gaussian:
\[
    E_{\mathcal{N}(0, \theta)}[X^n] =
    \begin{cases}
        (n-1)!! \,\theta^{\tfrac{n}{2}}, & n \text{ even}, \\[6pt]
        0, & n \text{ odd}.
    \end{cases}
\]

Hence, when $n$ is even,
\begin{align*}
    E[X^n] 
        &= (n-1)!! \int_\theta \theta^{\tfrac{n}{2}}
            \,\pi_{\text{hyper}}(\theta|r,\eta,\vartheta)\, d\theta \\[6pt]
        &= (n-1)!!\,E_{\pi_{\text{hyper}}(r,\eta,\vartheta)}[\theta^{\tfrac{n}{2}}] .
\end{align*}

The $k^{th}$ moment of a generalized gamma distribution is
\[
    E_{\pi_{\text{hyper}}(r,\eta,\vartheta)}[X^k]
    = \frac{ \vartheta^k \,\Gamma\!\left( \tfrac{\eta + 1.5 + k}{r} \right)}
           { \Gamma\!\left( \tfrac{\eta + 1.5}{r} \right)} .
\]

Substituting $k = \tfrac{n}{2}$, we obtain
\[
    E[X^n] =
    \begin{cases}
        \dfrac{ (n-1)!! \,\vartheta^{\tfrac{n}{2}}
        \Gamma\!\left(\tfrac{\eta + 1.5 + \tfrac{n}{2}}{r}\right) }
        { \Gamma\!\left(\tfrac{\eta + 1.5}{r}\right)}, & n \text{ even}, \\[12pt]
        0, & n \text{ odd}.
    \end{cases}
\]
\end{proof}

\section{Fit Categorization}
\label{app:fit_categorization}

We provide representative examples showing the different fit types in this appendix. Additionally, every fit and its classification can be found on \href{https://docs.google.com/spreadsheets/d/1LkjiR46a2qXNxUEQxDyO3FXKaUy_gMZwwjMA6BPMQ9g/edit?gid=1105455202#gid=1105455202}{this spreadsheet}.

\begin{figure}[H]
    \begin{center}
     \centering
     \includegraphics[width=.75\linewidth]{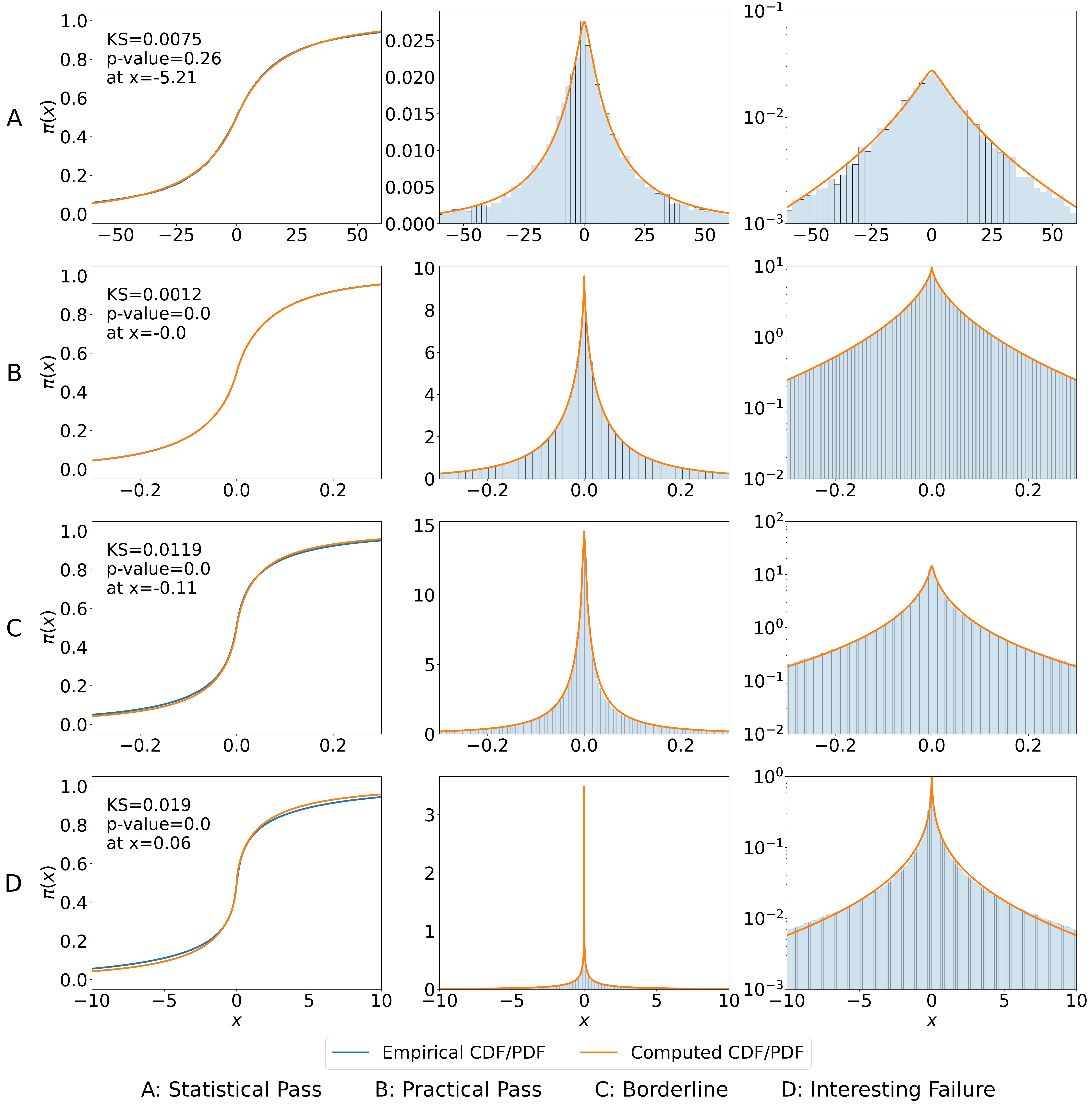}
  \end{center}
  \caption{This plot shows representative fit categorizations from all the non-trivial failure categories. The parameters for CDF, PDF and log PDF are provided in \Cref{tab:fit_categorization_non_trival_params} along with the datasets used for the empirical data.}
  \label{fig:fit_categorization_non_trival}
\end{figure}

\Cref{fig:fit_categorization_non_trival} shows categorizations from all of the non-trivial failure types. Panel A shows an example of a statistical pass. There is very little discrepancy when comparing the empirical and best-fit CDFs. When viewing the PDF plots, it's important to look at the histogram in addition to the KDE approximation. Oftentimes, for the better fitting curves, the computed PDF approximates the histogram even better than the KDE. Panel B highlights what it means to practically pass but not statistically pass. The fit in Panel B is almost 7 times better than Panel A when comparing the KS statistics. The reason that the fit in Panel B didn't statistically pass is the much larger number of samples, see \Cref{tab:fit_categorization_non_trival_params}. While there is not enough evidence to statistically pass, the better fit quality means that the model should perform better. Panel C shows a borderline asymmetric case. While it is quite slight, it is clear that the approximation error is due to the histogram being slightly asymmetric. This can be seen most clearly in the log KDE estimation. It can also be seen in the larger gap on the left tail of the CDF. This example was provided to show how strict practically passing was. This approximation may be sufficient for some applications; however, if there were any identifiable error, it would not be classified as a "passing" case. Panel D demonstrates the most common Interesting Failure case. This is when there is a large peak in the histogram. It is clear that the peak in the histogram is much smaller than the fitted PDF, and this error leads to undershooting in the tails. This case was deemed an interesting failure because it is not clear why the prior cannot model the shape.

\begin{figure}[H]
    \begin{center}
     \centering
     \includegraphics[width=.75\linewidth]{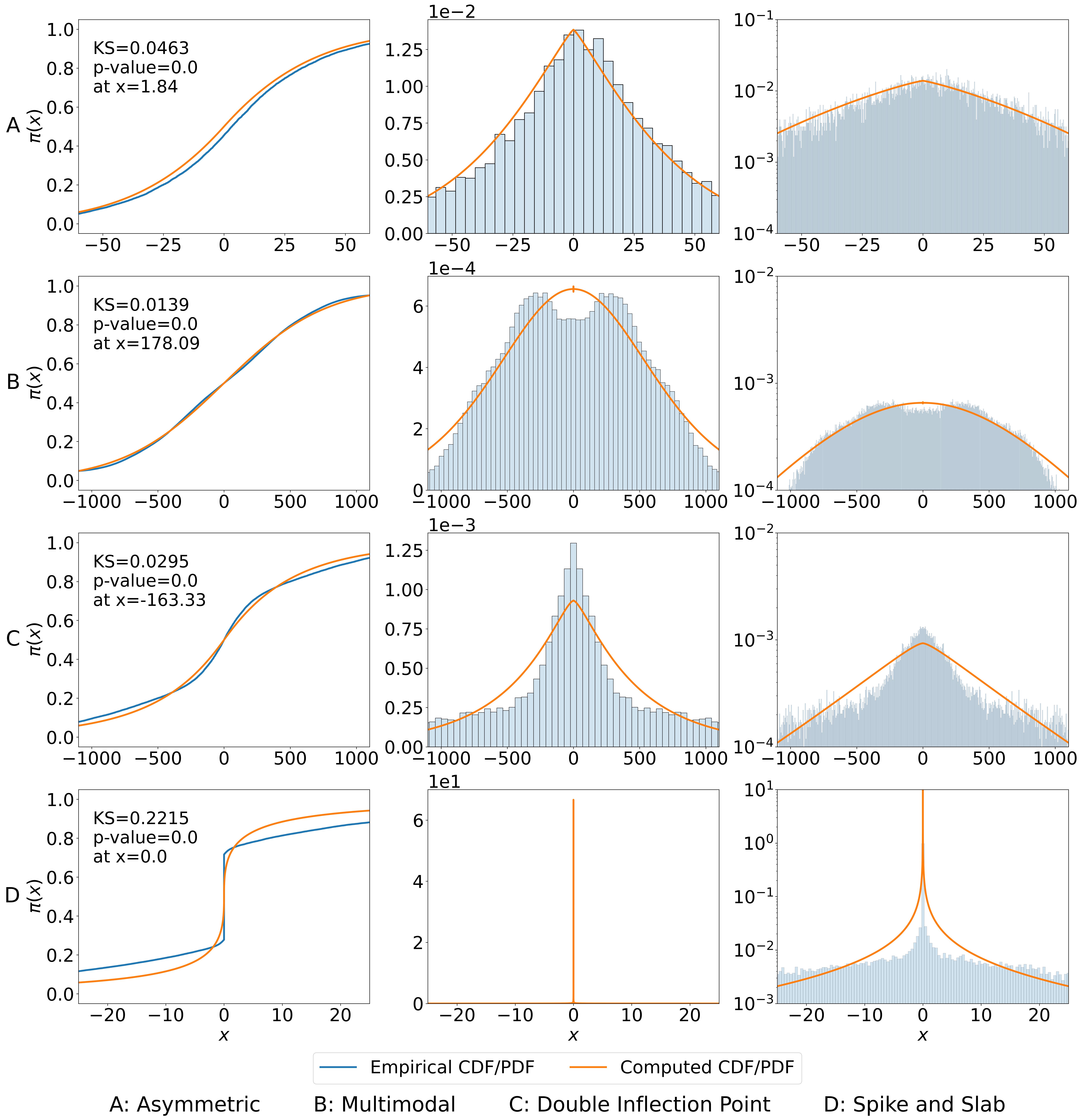}
  \end{center}
  \caption{This plot shows representative fit categorizations from all the trivial failure subcategories. The parameters for CDF, PDF and log PDF are provided in \Cref{tab:fit_categorization_trivial_failure_params} along with the datasets used for the empirical data.}
  \label{fig:fit_categorization_trivial_failure}
\end{figure}

It is clear that no set of parameters can fit the trivial failure cases displayed in \Cref{fig:fit_categorization_trivial_failure}. All of these qualities cannot be modeled by our prior model. Panel A shows an asymmetric distribution. While this case was obvious from all three plots, closer cases were determined by looking if the CDF was higher or lower on both sides of zero. Because the distribution should be symmetric, being higher on one side should correlate with being lower on the other. Panel B shows a Multimodal distribution with two modes. Some of these cases may be caused by applying the symmetrization procedure described in Sec \ref{section:normalization} to a distribution not centered around $0$. Other cases truly had multiple modes. Panel C shows a double infection point, which cannot be modeled by the distribution. These cases tended to be rarer, and we are not 100\% sure what caused them. Panel D shows a severe Spike and Slab distribution. Because this only happened to the middle to high layers of the spaceNet dataset under the Haar wavelet transform, we believe it may be caused by pictures that are all or mostly water.

\begin{table}[ht]
\centering
\caption{Best fit parameters and dataset details for Panels A--D.}
\label{tab:fit_categorization_non_trival_params}
\begin{tabular}{c c c c c c c c}
\hline
Panel & $r$ & $\eta$ & $\vartheta$ & Dataset & Transform/Filter & Channel & Samples \\
\hline
A & $0.09$ & $1.9$ & \num{1.743e-15} & agiriVision & Wavelet, diagonal detail, layer 3 & Green & $18{,}000$ \\
B & $0.29$ & $-0.6$ & \num{2.792e-4} & segmentAnything & Wavelet, diagonal detail, layer 10 & Blue & $463{,}470{,}592$ \\
C & $0.08$ & $0.4$ & \num{3.272e-20} & coco (indoor) & Wavelet, vertical detail, layer 9 & Green & $26{,}279{,}936$ \\
D & $0.30$ & $-1.07$ & \num{2.927e0} & spaceNet & learned filters, filter 49 & --- & $480{,}000{,}000$ \\
\hline
\end{tabular}
\end{table}

\begin{table}[ht]
\centering
\caption{Representative fit categorizations from trivial failure categories.}
\label{tab:fit_categorization_trivial_failure_params}
\begin{tabular}{c c c c c c c c}
\hline
Panel & $r$ & $\eta$ & $\vartheta$ & Dataset & Transform/Filter & Channel/Slice & Samples \\
\hline
A & $8.3$ & $-0.4$ & \num{2.938e-3} & coco (outdoor) & Wavelet, horizontal detail, layer 3 & Gray & $9{,}784$ \\
B & $0.2$ & $19$ & \num{3.574e-5} & MRI 2D & Wavelet, diagonal detail, layer 4 & Axial slice & $180{,}942$ \\
C & $0.55$ & $-0.1$ & \num{7.892e5} & MRI 3D & Wavelet, DAD detail, layer 4 & --- & $22{,}538$ \\
D & $0.3$ & $-1.3$ & \num{2.706e2} & spaceNet & Wavelet, diagonal detail, layer 3 & Blue & $54{,}416$ \\
\hline
\end{tabular}
\end{table}

\section{CDF Quadrature Routine}
\label{app:CDF_quadrature}

Based on equation \eqref{eqn:prior}, evaluating the CDF of our prior at some $x$ means we must evaluate the following integral:
\begin{equation}
    F(T|r,\eta,\vartheta) = \int_{x = -\infty}^T \int_{\theta = 0}^\infty \,\pi_{\mathcal{N}(0, \theta)}(x|\theta) 
            \,\pi_{\text{hyper}}(\theta|r,\eta,\vartheta)\, d\theta \, dx
    \label{eq:cdf_equation}
\end{equation}

Applying \Cref{eq:cdf_equation} for each KS test would be far to slow due to the number of CDFs and large sample size. Instead, each CDF was approximated. First, we choose a range of points that spanned a majority of the distribution's support using Chebyshev's Bound. Then we evaluate the CDF at these points and fit a cubic spline that approximates the true CDF. Additionally, to increase the accuracy in the $\eta < 0$ region where the distribution puts a lot of mass near $0$, $x$ values are recursively placed if their neighboring values have a difference in $\text{CDF} > 0.02$. The final approximation is tested using a KS test against samples generated from the prior to ensure it is a faithful approximation.

Numerically evaluating \eqref{eq:cdf_equation} using a double quadrature scheme proved to be difficult. In order to make it more accurate, we can use the CDF of the Generalized Gamma Distribution. We can do this by switching the order of integration using a change in variables. Recall that $\pi_{\text{hyper}}$ is a generalized gamma distribution.

Let $z = \frac{x}{\sqrt{\theta}}, \theta = \theta$:
\begin{align}
     F(T|r,\eta,\vartheta) &= \int_{x = -\infty}^T \int_{\theta = 0}^\infty \,\pi_{\mathcal{N}(0, \theta)}(x|\theta) 
            \,\pi_{\text{hyper}}(\theta|r,\eta,\vartheta)\, d\theta \, dx \nonumber \\[6pt]        
    &= \int_{x = -\infty}^T \int_{\theta = 0}^\infty \frac{1}{\sqrt{2\pi}\theta} \exp{\left(\frac{x^2}{2\theta}\right)} 
        \pi_{\text{hyper}}(\theta|r,\eta,\vartheta)\, d\theta \, dx \nonumber\\[6pt]     
    &= \int_{\theta = 0}^\infty \int_{z = -\infty}^{\frac{T}{\sqrt{\theta}}}  \frac{1}{\sqrt{2\pi}} \exp{\left(\frac{z^2}{2}\right)} 
        \pi_{\text{hyper}}(\theta|r,\eta,\vartheta)\,dz \, d\theta  \nonumber \\[6pt]
     &= \int_{\theta = 0}^\infty \int_{z = -\infty}^{\frac{T}{\sqrt{\theta}}}  \pi_{\mathcal{N}(0, 1)}(z|\theta) 
            \,\pi_{\text{hyper}}(\theta|r,\eta,\vartheta) \, dz \, d\theta \label{eq:cdf_integral_theta_then_z}
\end{align}

\begin{figure}[H]
    \begin{center}
     \centering
     \includegraphics[width=.5\linewidth]{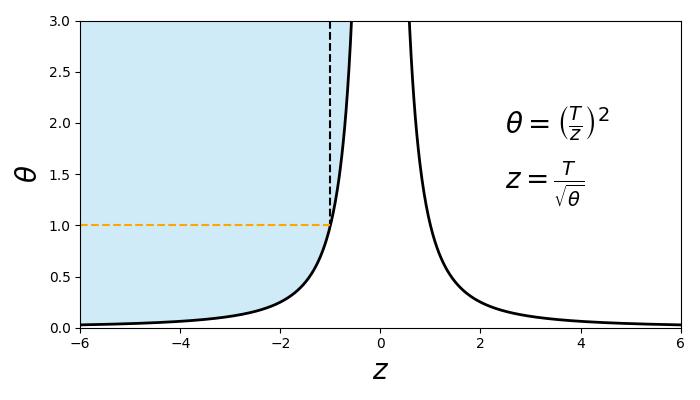}
  \end{center}
  \caption{The blue shaded region shows the area being integrated over equations \eqref{eq:cdf_integral_theta_then_z}, \eqref{eq:cdf_integral_z_then_theta}.
  When evaluating the inner integral in equation \eqref{eq:cdf_integral_z_then_theta} at a fixed $\theta$ (shown $\theta= 1$), the orange line indicates that the bounds are from $-\infty$ to the curve $z = \frac{T}{\sqrt{}\theta}$. The outer integral bounds go from $0$ to $\infty$. After we switch the order of integration, we evaluate the inner integral at a fixed $z$ (shown $z = 1)$. The black line indicates that the lower bound is the function $\theta = \left(\frac{T}{z}\right)^2$ and the upper bound is $\infty$. Because $T,z$ share the same sign, and $T \leq 0$, the outer bound is from  $-\infty$ to $0$.}
\label{fig:integration_parameter_space_map}
\end{figure}

Because the distribution is symmetric around $0$, we only need to consider the case $T\leq0$. If $T > 0$, $F(T|r,\eta,\vartheta) = 1 - F(T|r,\eta,\vartheta)$. Looking at \Cref{fig:integration_parameter_space_map} and keeping in mind that the function $z = \frac{T}{\sqrt{\theta}} \implies \theta = \left(\frac{T}{z}\right)^2$ we can switch the order of the integral again.

\begin{align}
     F(T|r,\eta,\vartheta)  &= \int_{z = -\infty}^0 \int_{\theta = \left(\frac{T}{z}\right)^2}^{\infty}  \pi_{\mathcal{N}(0, 1)}(z|\theta) 
            \,\pi_{\text{hyper}}(\theta|r,\eta,\vartheta)\, d\theta \, dz \label{eq:cdf_integral_z_then_theta}\\[6pt]
    &= \int_{z = -\infty}^0 \pi_{\mathcal{N}(0, 1)}(z|\theta) \left(\int_{\theta = \left(\frac{T}{z}\right)^2}^{\infty}   
            \,\pi_{\text{hyper}}(\theta|r,\eta,\vartheta)\, d\theta \, \right)dz \nonumber \\[6pt]
     &= \int_{z = -\infty}^0 \pi_{\mathcal{N}(0, 1)}(z|\theta) \left(1 - F_{\pi_{\text{hyper}}(r,\eta,\vartheta)}\left(\frac{T}{z}\right)^2 \right)dz \nonumber
\end{align}

$F_{\pi_{\text{hyper}}(r,\eta,\vartheta)}$ is the generalized gamma CDF which can be computed in closed form in terms of the $\Gamma$ function.

\end{document}